\documentclass[12pt,preprint]{aastex}

\slugcomment{To be submitted to \apj .}

\shorttitle{Searching for Planets Around Vega and $\epsilon$ Eri}
\shortauthors{Heinze and Hinz}

\begin{document}

\title{Deep L' and M-band Imaging for Planets Around Vega and $\epsilon$
  Eridani\altaffilmark{1}}

\author{A. N. Heinze}
\affil{Swarthmore College, 500 College Avenue, Swarthmore, PA 19081}
\email{aheinze1@swarthmore.edu}

\author{Philip M. Hinz}
\affil{Steward Observatory, University of Arizona, 933 N Cherry Avenue, Tucson, AZ 85721-0065}
\email{phinz@as.arizona.edu}

\author{Matthew Kenworthy}
\affil{Steward Observatory, University of Arizona, 933 N Cherry Avenue, Tucson, AZ 85721-0065}
\email{mkenworthy@as.arizona.edu}

\author{Douglas Miller}
\affil{Steward Observatory, University of Arizona, 933 N Cherry Avenue, Tucson, AZ 85721-0065}
\email{dlmiller@as.arizona.edu}

\author{Suresh Sivanandam}
\affil{Steward Observatory, University of Arizona, 933 N Cherry Avenue, Tucson, AZ 85721-0065}
\email{suresh@as.arizona.edu}

\altaffiltext{1}{Observations reported here were obtained at the MMT
Observatory, a joint facility of the University of Arizona and the
Smithsonian Institution.}

\begin{abstract}
We have obtained deep Adaptive Optics (AO) images of
Vega and $\epsilon$ Eri to search for planetary-mass
companions.  We observed at the MMT 
in the $L'$ (3.8 $\mu$m) and $M$ (4.8 $\mu$m) bands using Clio, 
a recently commissioned imager optimized for these wavelengths.  
Observing at these long wavelengths represents a departure from
the $H$ band (1.65 $\mu$m) more commonly used for AO imaging
searches for extrasolar planets.  The long wavelengths offer
better predicted planet/star flux ratios and cleaner (higher Strehl) AO images,
at the cost of lower diffraction limited resolution and higher sky background.
We have not detected any planets or planet candidates around Vega or
$\epsilon$ Eri.  We report the sensitivities obtained around
both stars, which correspond to upper limits on any planetary companions
which may exist.  The sensitivities of our $L'$ and $M$ band
observations are comparable to those of the best $H$-regime observations
of these stars.  For $\epsilon$ Eri our $M$ band 
observations deliver considerably better sensitivity to close-in planets
than any previously published results, and we show that the $M$ band
is by far the best wavelength choice for attempts at ground-based
AO imaging of the known planet $\epsilon$ Eri b.  The Clio camera 
itself with MMTAO may be capable of detecting $\epsilon$ Eri b 
at its 2010 apastron, given a multi-night observing campaign.  Clio appears
to be the only currently existing AO imager that has a realistic
possibility of detecting $\epsilon$ Eri b.
\end{abstract}


\section{Introduction}
Early space based observations with the IRAS satellite identified four bright,
nearby stars with strong IR excesses: $\beta$ Pic, Vega, Fomalhaut, and
$\epsilon$ Eri \citep{IRASv1,IRASa1,IRASa2,IRASa3}.  The only reasonable 
explanation for these excesses is that
the systems contain substantial dust, which is warmed by starlight until
it radiates brightly in the IR because of the large total surface 
area of its numerous small grains (see for example \citet{backman, HD4796, deller}).

The dust in these systems cannot be primordial but must be 
continually generated by the grinding down of larger bodies such as 
asteroids \citep{backman, HD4796, deller}.
The stars are therefore said to have `debris disks'.  The clear implication
is that each of these stars has at least an asteroid belt, and probably
a more extensive planetary system, because it is unlikely that an asteroid belt
would form without planets also forming, or that it would continue 
to grind down without ongoing gravitational stirring due to planets.

Theoretical models (e.g. \citet{bur} and \citet{bar}) predict that 
it should be possible to make direct
images of giant planets orbiting nearby, young stars,
using the current generation of large ground-based telescopes with
adaptive optics (AO).  These observations are only possible at near infrared
wavelengths from about 1-5 $\mu$m, where giant planets are self-luminous due
to the gravitational energy converted to internal heat in their formation
and subsequent slow contraction.  Because giant planets radiate this
energy away over time, they become cooler and fainter as they age.  
The youngest nearby stars are therefore the most promising targets for AO
surveys attempting to image self-luminous giant planets.  

Each of the four debris-disk stars discovered using IRAS is relatively
young, so orbiting giant planets might be detectable if any exist.  We
have imaged the two stars most easily observable from Northern Hemisphere
sites: Vega and $\epsilon$ Eridani.  Vega's age is about 0.3 Gyr
\citep{agevega}, while the age of $\epsilon$ Eri is about 
0.56 Gyr \citep{fischer}.
Besides the dust-dispersion timescale
argument mentioned above for the existence of planetary systems around
these stars, asymmetries in the dust distributions around each have led to hypotheses
that the dust is being gravitationally sculpted by giant planets orbiting
at large distances \citep{ozernoy,quillen,wyatt,wilner,deller,marsh}.
In the case of Vega there are suggestions that the dust may reveal the mass
and approximate position of a giant planet \citep{wilner,deller}.  For 
$\epsilon$ Eridani, in addition to evidence for a planet in a distant
orbit that may be sculpting the dust \citep{deller,benedict}, there is the
the radial velocity and astrometric detection of the closer-in planet
$\epsilon$ Eri b \citep{benedict}.  The combination of radial 
velocity and astrometry
observations permits a full orbital solution yielding ephemerides 
for the separation and position angle of $\epsilon$ Eri b \citep{benedict},
making this the most promising case yet where attempts to image a 
known extrasolar planet can target a specific location.

Most imaging searches for extrasolar planets to date have used either the
$H$ band (1.5 - 1.8 $\mu$m) or other filters in the same wavelength regime (see for
example \citet{hsttwa7a, masciadri, sdi1, epsindia, biller, GDPS}).
The magnitude vs. mass tables of \citet{bar} and the theoretical
spectra of \citet{bur} show clearly why the $H$ band is usually chosen:
giant planets are predicted to be very bright at these wavelengths, much brighter than black bodies
at their effective temperatures.  Detector formats are large, technology well 
developed, and sky backgrounds faint at the $H$ band relative to longer wavelengths.

However, theoretical models indicate that planet/star flux ratios are much more
favorable at the longer wavelength $L'$ and $M$ bands (3.4-4.1 $\mu$m and 4.5-5.0 $\mu$m, respectively).
For planets at sufficiently large separations, or planets orbiting faint stars, the
planet/star flux ratio is not relevant.  Rather, it is the planet's brightness relative
to the sky background and/or detector read noise that matters.  In this regime
the very high sky background in the $L'$ and $M$ bands prevents them from being
as sensitive as the $H$ band regime.  However, close to very bright
stars the background becomes irrelevant and only the planet/star flux ratio
matters.  Under these circumstances using the longer wavelengths makes sense.

Vega is a magnitude 0.0 standard star and is among the brightest stars in the sky at
almost any wavelength.  $\epsilon$ Eri, while not impressive at visible wavelengths, is
a very bright magnitude 1.9 at $H$ band.  The stars are therefore excellent targets
for Clio, an $L'$ and $M$ band optimized AO camera that had its first light
on the MMT in June 2005 \citep{oldvega}.  We have made deep $\sim$1 hour
integrations in both the $L'$ and $M$ bands on both stars.  Our $M$-band 
observations are the deepest ground-based images yet published in this band.

In Section \ref{obs} we present our observations and data analysis strategy.
In Section \ref{sensanal}, we describe our methods of analyzing
our sensitivity, and present our sensitivity results.  Blind sensitivity tests
in which simulated planet images were inserted directly into the raw data show that
we have obtained 100\% completeness for sources at 10$\sigma$ signficance, 77\% completeness for 7$\sigma$ sources, and 41\% 
completeness for 5$\sigma$ sources, where $\sigma$ is an estimate of the
RMS noise amplitude in the image at the spatial scale of the PSF core (the
relevant scale for detection of faint point sources).  We note
that no other planet-imaging papers to date present such careful blind tests
in their sensitivity analyses, and that the fact that our tests result in somewhat
lower completeness values at each significance level than might have been expected
suggests such tests should always
be attempted and may result in a need to revise some sensitivity estimates to more
conservative values.

In Section \ref{vega} we compare the sensitivity we have obtained around
Vega to that of other deep
observations of Vega, and to the expected brightness of planets that have been hypothesized to explain the
dust distribution.  In Section \ref{eri}, we present the same comparisons for
$\epsilon$ Eri, and in Section \ref{conclusion} we present the conclusions of our
study.

\section{Observations and Data Analysis} \label{obs}

\subsection{The Instrument} \label{clio_ins}

The Clio instrument we used for our observations has been well described 
elsewhere (\citet{freed},
\citet{suresh}, and \citet{oldvega}).  We present only a brief overview here.

The MMT AO system delivers a lower thermal background than other AO systems because it
uses the world's first deformable secondary mirror, thereby avoiding the multiple warm-mirror
reflections (each adding to the thermal background) that are needed in AO systems 
where the deformable mirror is not the secondary.  This unique property makes the 
MMT ideal for AO observations in wavelengths such as the $L'$ and $M$ bands
that are strongly affected by thermal glow.  Clio was developed to take advantage of
this to search for planets in these bands.  It saw first light as a simple
imager offering F/20 and F/35 modes.  The design allowed for coronagrapic
capability, which has since been developed \citep{phaseplate} but was not fully operational
at the time of our Vega and $\epsilon$ Eri observations.  In the F/20 mode
we used for all the observations of Vega and $\epsilon$ Eri, Clio's 
field of view is 15.5$\times$12.4 arcseconds.  Its plate scale 
is $0.04857 \pm 0.00003$ arcseconds per pixel, which gives finer than Nyquist
sampling of the diffraction-limited PSF of the MMT in the $L'$ and $M$ bands.

\subsection{Observing Strategy} \label{obsstrat}  
We carry out $L'$ and $M$ band imaging with Clio using
the technique of nod-subtraction, in which we take images of our 
target star in two different
telescope positions offset typically by about 5.5 arcsec, and then 
subtract the images taken in one position from those taken in the 
other to remove artifacts from the bright sky background and 
detector imperfections. Since the star is present on images taken 
in both positions, both provide useful
science data.  Nod-subtraction does result in a dark negative image of the star
reducing the sensitivity in part of each image, but the area affected is
fractionally small, far (5.5 arcsec) from the star, where planets are less
likely to be found, and can be placed away from objects of potential interest
by a good choice of the nod direction.  We also have alternative ways of
processing nodded data that do away with the dark images entirely.

We typically nod the telescope every 2-5 minutes, which appears to be fast enough
that variations in the sky background are sampled well enough to be
essentially removed.  We take 5 or 10 images in each nod position, 
each of which typically represents
about 20 seconds worth of data. A full data set consists of 100-500
such images.

We choose the exposure for most of the images so that the sky background level
is about 70\% of the detector full-well.  At such exposure times the cores
of bright stars such as Vega and $\epsilon$ Eri are saturated, but
optimal sensitivity is obtained to faint point sources beyond the
saturation radii.  When possible, we interleave a few nod cycles of
shorter exposures yielding unsaturated star images into the sequence
of longer exposure images.  This allows us to measure the unsaturated 
PSF under the exact conditions of a particular observing sequence.  
We achieved the PSF measurement with $\epsilon$ Eri, but
Vega proved too bright for us reasonably to obtain unsaturated images.  We
used other stars observed close in time to our Vega observations to
provide a reference PSF for the Vega data.

Tables \ref{tab:vobs01} and \ref{tab:vobs02} give details of our observations.
The June 2006 $M$ band Vega observations had
far higher sky noise than the April 2006 data, possibly because of the higher
thermal background during warm summer weather, and therefore were not used in 
calculating the final sensitivity.

\clearpage
\begin{deluxetable}{llcrrr}
\tablewidth{0pt}
\tablecolumns{6}
\tablecaption{Observations of Science Targets: Basic Parameters \label{tab:vobs01}}
\tablehead{\colhead{Star} & \colhead{Date Obs} & \colhead{Band} & \colhead{Clio int(msec)} & \colhead{Coadds} & \colhead{\# Images}} 
\startdata
Vega & April 12, 2006 & $L'$ & 2000 & 10 & 160 \\
Vega & April 13, 2006 & $M$ & 200 & 90 & 110 \\
Vega & June 10, 2006 & $M$ & 100 & 50 & 558 \\
Vega & June 11, 2006 & $M$ & 120 & 100 & 180 \\
$\epsilon$ Eri & September 09, 2006 & $M$ & 130 & 100 & 180\\
$\epsilon$ Eri & September 11, 2006 & $L'$ & 1500 & 15 & 184 \\
\enddata
\tablecomments{Clio int(msec) refers to the nominal single-frame
exposure time in Clio.  The integrate-while-reading mode used
in high efficiency science imaging causes the true single-frame
exposure time to be about 59.6 msec longer than the nominal exposures
listed here.  Coadds is the number of frames internally coadded
by Clio to produce a single 2-D FITS image.}
\end{deluxetable}

\begin{deluxetable}{lccccr}
\tablewidth{0pc}
\tablecolumns{5}
\tablecaption{Observations of Science Targets: Data Acquired \label{tab:vobs02}}
\tablehead{\colhead{Star} & \colhead{Date} & \colhead{Band} & \colhead{Exposure(sec)} & \colhead{Mean Airmass} & \colhead{Rotation}}
\startdata
Vega & April 12, 2006 & L' & 3295.4 & 1.018 & $80.63^{\circ}$ \\
Vega & April 13, 2006 & M & 2570.0 & 1.026 & $36.39^{\circ}$ \\
Vega & June 10, 2006 & M & 4452.8 & 1.034 & $72.36^{\circ}$ \\
Vega & June 11, 2006 & M & 3232.8 & 1.054 & $25.53^{\circ}$ \\
$\epsilon$ Eri & September 09, 2006 & M & 3412.8 & 1.334 & $23.41^{\circ}$ \\
$\epsilon$ Eri & September 11, 2006 & L' & 4304.5 & 1.342 & $36.92^{\circ}$ \\
\enddata
\tablecomments{The observations in June were plagued with high
sky noise, which may have been due to the higher thermal background
during warm summer weather.  Adding them to the
April M-band data on Vega did not significantly increase the
sensitivity to objects far from the star, though in the speckle
dominated regime near the star, the sensitivity did increase
by about 40 \%.}
\end{deluxetable}
\clearpage

\subsection{Data Analysis} \label{dataan}
\begin{figure*}
\plotone{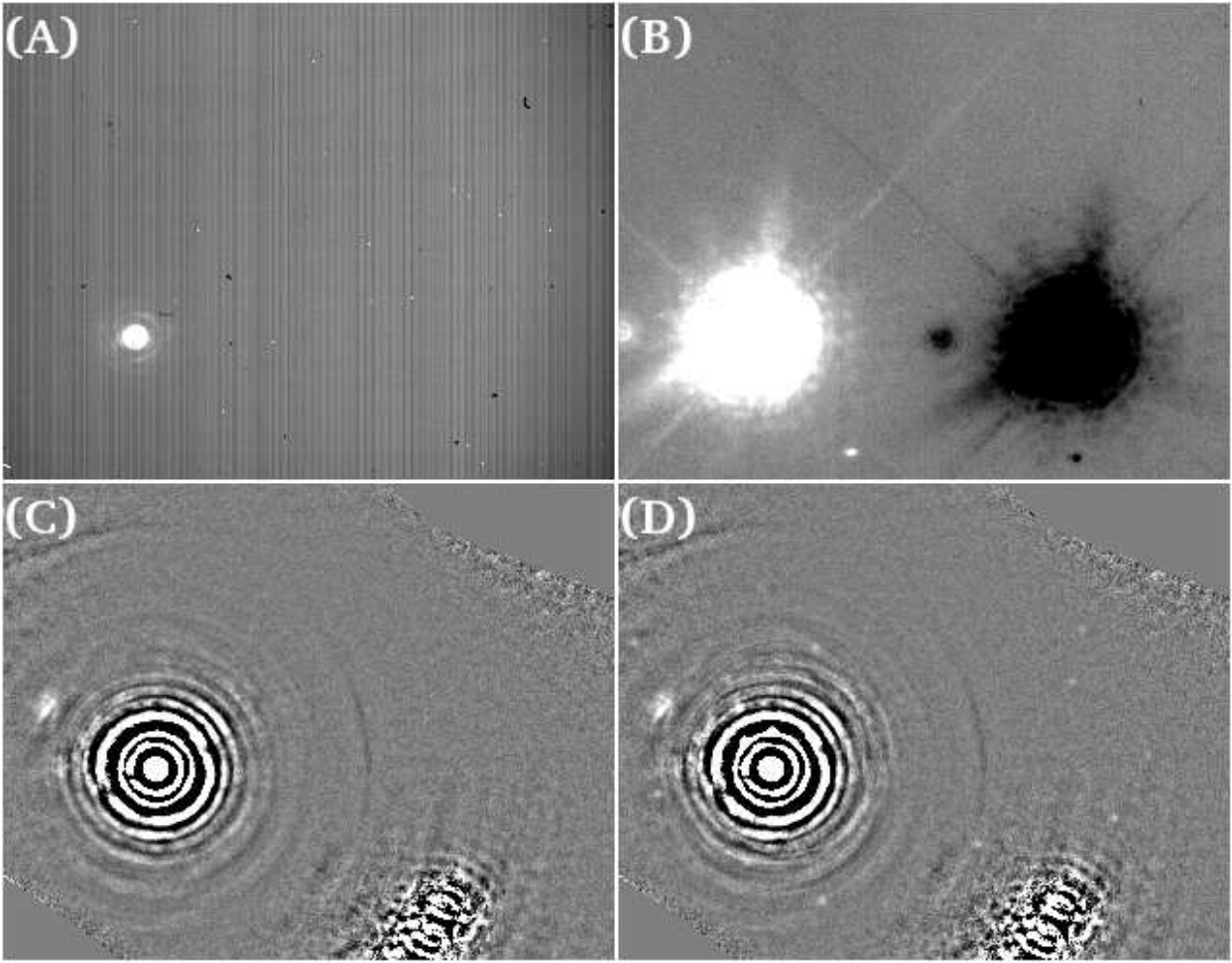} 
\caption[Examples of our Vega $M$ Band Images] {(A) 
Raw single $M$ band image of Vega.  (B) Nod subtracted, 
processed version of the same image just before shift and rotation.
Contrast stretched 100 times more than in Panel A. (C) Final master
$M$ band image of Vega, consisting of 110 Panel B-like
images, shifted, rotated, and coadded.  Contrast stretched 10 times more
than in Panel B.  (D) Like Panel C, but with fake planets added
to the raw data.  The field shown in all panels is about 15.5$\times$ 12.5
arcseconds.  The unsharp masking which has removed the bright 
stellar halo in Panels C and D is responsible for
the black spaces between the inner diffraction rings.
The noisy region at bottom right in these panels is due
to the negative stellar images from nod subtraction.  
}
\label{fig:tileim}
\end{figure*}

Our Clio image processing pipeline
will be described in more detail in a future paper.  
Here we briefly state that our baseline processing involves dark
subtraction; flat fielding; nod subtraction; several iterations
of different types of deviant (`hot') pixel removal; a pattern
noise correction (Figure \ref{fig:tileim} Panel B shows
an example image at this stage); shifting, rotation, and 
zeropadding in a single
bicubic spline operation; final stacking; and then unsharp masking 
of the stacked image using a gaussian kernel 3-4 times wider than 
the PSF.

For the final image stacks we use a creeping mean algorithm with 
20\% rejection.  This algorithm works by finding the
mean of all values for a given pixel, rejecting the most deviant one,
finding the new mean, rejecting the new maximally deviant value,
etc, until the specified rejection fraction is reached.  For data sets
where ghosts or other artifacts can render a large fraction of the
data at a given location deviant, the creeping mean produces a cleaner
final stack than the median.  Figure \ref{fig:tileim} Panel A shows
an example of a raw image; Panel B shows a partially processed version
of the same image just before shifting and rotation;
and Panels C and D show examples of final stacked images after
unsharp masking.

In addition to the image made using our baseline processing, 
we make images using two types of more advanced processing, 
one that avoids the negative star
images from standard nod subtraction at the cost of slightly
increased noise, and one that includes
subtraction of the stellar PSF using a technique similar
to the angular differential imaging (ADI) described by \citet{marois}.
We use all three images when we search for companions, since
the detection of a faint companion on images processed in more
than one way increases the likelihood that it is real.  We also
construct a separate sensitivity map for each of the three 
differently-processed
master images, and then combine them into a single master sensitivity
map.  Since the different processing methods obtain optimal sensitivity
at different locations, we set the sensitivity at a given location
on the master map to the best sensitivity obtained at that location
on any of the three separate maps.  Details on how the separate
sensitivity maps themselves are made may be found in 
Section \ref{sensanal}.

\clearpage

Intensive image processing such as we describe here is often
used for AO planet search data, where high contrast
is required and artifacts must be aggressively removed. 
Such processing can remove flux from the faint point sources
whose detection it is intended to facilitate.  Careful tests of
our processing methods, however, indicate that the flux loss 
from a faint PSF is no more than about 10\%, and appears to be close 
to zero in most cases.

\section{Sensitivity Measurements and Source Detection Tests} \label{sensanal}

\subsection{Sensitivity Estimation}
We create a sensitivity map from each stacked master image produced
by the processing outlined above.  We are careful to measure the
noise at the relevant spatial scale -- that is, the scale of the
PSF.  Our method requires an unsaturated star image taken under
similar conditions to the science data, and therefore representing
a good estimate of the PSF.  We perform a two-parameter least
square fit centered on each pixel in the master science image
in turn, with the two parameters being the amplitude of a PSF
centered on that pixel, and a constant background value.  This
fit is performed within a disk of six pixel radius about each given
pixel.  The best-fit PSF amplitude from the fit
centered on each pixel of the master science image becomes the
value of the corresponding pixel of a new image: 
the PSF amplitude map. This PSF amplitude
map image is essentially the result of PSF-fitting photometry 
centered in turn on every pixel in the original master science image.  
This PSF-fitting has, of course, mostly measured simply noise -- the 
point is that it has measured the noise at the spatial scale of the PSF.

Our method may be expected to produce results somewhat similar to
the `matched filter' technique (see for example \citet{matchfilt}),
although the least-square fitting that we use is mathematically
more sophisticated than the straightforward convolution used in
a matched filter.  The most obvious advantage of our method is that
it automatically fits and removes any slowly-varying background (since
our least square fit determines a separate background value within 
the disk centered on each pixel), while an ordinary matched filter 
requires the separate construction of a background model.

The noise in the amplitude map constructed by our PSF-fitting
accurately reflects the PSF-scale noise in the original image -- that
is, the noise at the spatial frequencies relevant for the 
detection of real point sources.  We calculate the
sensitivity at every point in the original image by computing the RMS
in an 8 pixel radius aperture about that point on the PSF amplitude map
(for regions too close to the star, where a circular aperture would not produce
accurate results, we use a 45-pixel long arc at constant radius from the
star instead).  Note that since the data are only slightly oversampled,
both the 8 pixel radius disk and the 45-pixel long arc span many resolution
elements or speckles.  Calculating the RMS on the PSF amplitude map
rather than the original master image takes into account
spatial correlations between pixels (that is, the fact that the
noise in adjacent pixels is not independent).  This is a
large effect in the case of speckle noise.  

We note that many previous planet-imaging papers have not
used a sensitivity estimator mathematically able to account
for correlated noise in speckles --- or, at least, have not devoted 
sufficient space to the description of their sensitivity
estimator to make it clear whether or not it properly
measures correlated noise.  Estimators that contain an implicit
mathematical assumption that the noise is independent from
one pixel to the next can significantly overestimate the
sensitivity in speckle-dominated regions close to the star.
The careful design, description, and testing of sensitivity estimators is
an important task, because in the case of a non-detection all 
the science rests on upper limits set through sensitivity estimation.  
The only observational planet-imaging paper we are aware of prior to
this work in which a sensitivity estimator able to account for
correlated noise is clearly described is \citet{GDPS}. (However,
we may safely assume that \citet{marois} used the same estimator
as \citet{GDPS}. \citet{hinkley} also used, and carefully 
described, such an estimator in their paper to set limits on
brown dwarfs in close orbits around Vega.)

\subsection{Testing the Sensitivity Estimator}

To test the accuracy of our sensitivity estimator, we conducted blind
tests in which fake planets were inserted into the raw
data.  The altered images were then processed in exactly the same way
as the original raw data, and the `planets' were
detected using both automatic and manual methods by an experimenter
who knew neither their positions nor their number.  These planets were inserted
at fixed nominal significance levels of 10$\sigma$, 7$\sigma$, and 5$\sigma$ based
on the sensitivity map.  We conducted such tests for each of
our four data sets (the $L'$ and $M$ band data sets for each of the two
stars).  The final result of each test was that every inserted planet
was classified as `Confirmed', `Noticed', or `Unnoticed'.
`Confirmed' means the source was confidently detected and
would certainly be worthy of long-exposure followup observations at the MMT.
If a source is detected with this confidence level in an unaltered data set,
there is no significant doubt it is a real object.  In calculating our
completeness, we count only confirmed sources as true detections.
`Noticed' means the source was flagged by our automatic detection algorithm,
or noticed as a possible real object during the purely manual phase
of planet-searching, but could not be confirmed beyond reasonable doubt.
Many spurious sources are `Noticed' whereas the false-positive rate for
`Confirmed' detections is extremely low, with none for any of the data
sets discussed here.  `Unnoticed' means a fake planet was not automatically
flagged or noticed manually.  

The end result of the four blind sensitivity tests was that at 10$\sigma$, 50 of
50 total inserted planets were confirmed, giving us 100\% completeness
to the limits of the statistical accuracy of the test.  At 7$\sigma$,
23 of 30 total inserted sources were confirmed, giving us 77\% completeness,
and 29 of the 30 sources were at least noticed.
At 5$\sigma$, 11 of 27 total inserted planets were confirmed, for 41\%
completeness, and 23 of the 27 sources were at least noticed.  
In addition to the completeness levels for confirmed sources, the percentages 
of fake planets that were at least noticed is of potential
interest for setting limits: 100\% of 10$\sigma$ sources, 97\% of 7$\sigma$
sources, and 85\% of 5$\sigma$ sources were at least noticed.
We note that if we had quoted 5$\sigma$ sensitivities without conducting a
blind sensitivity test we would have significantly overestimated our true
high-completeness sensitivity.  Most papers in the field of 
planet-imaging surveys do in
fact quote 5$\sigma$ limits, and do not verify their validity by
a blind test.  

In our sensitivity experiments there were no false positives among 
the `Confirmed' sources.  Many spurious sources were classified as
`Noticed', which is why we do not count `Noticed' sources as detections
for completeness purposes.  The conclusion of our fake planet experiments
is that our detection strategy has an extremely low
false alarm probability, and delivers the completeness values given above.  
The fact that a large majority of low significance sources were noticed,
even if not confirmed, indicates that upper limits stronger than those
implied by our formal completeness values may be set on planets in clean 
regions of an image where no spurious sources were noticed.

\subsection{Final Sensitivity Results} \label{finsens}
\begin{figure*}
\plotone{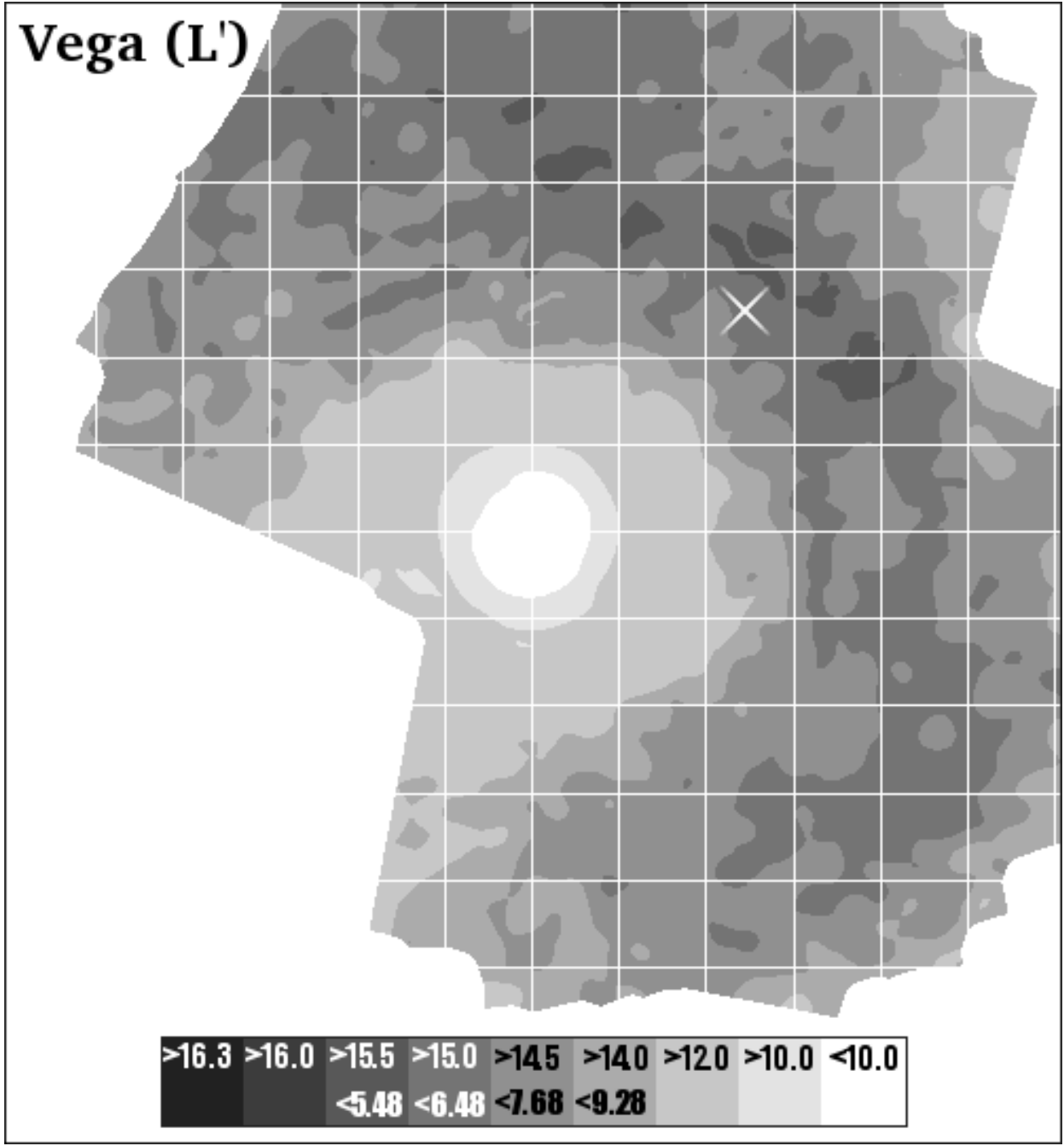} 
\caption[Sensitivity contour map for the Vega $L'$ observations] {
10$\sigma$ sensitivity contour map for our Vega $L'$ observations in magnitudes.
The grid squares superposed on the figure for astrometric reference are 2x2 arcsec.
The approximate location of the hypothetical planet from \citet{wilner}
is marked with a white `X'.  The best areas in this image give sensitivity
to objects fainter than $L'$ = 15.5.  The numbers at the top of the colorbar
give the sensitivity of each contour in magnitudes, while the numbers at
the bottom give the equivalent value in MJ, where applicable, based on the
models of \citet{bur} with the age set to 0.3 Gyr.
}
\label{fig:vegamap}
\end{figure*}

\begin{figure*}
\plotone{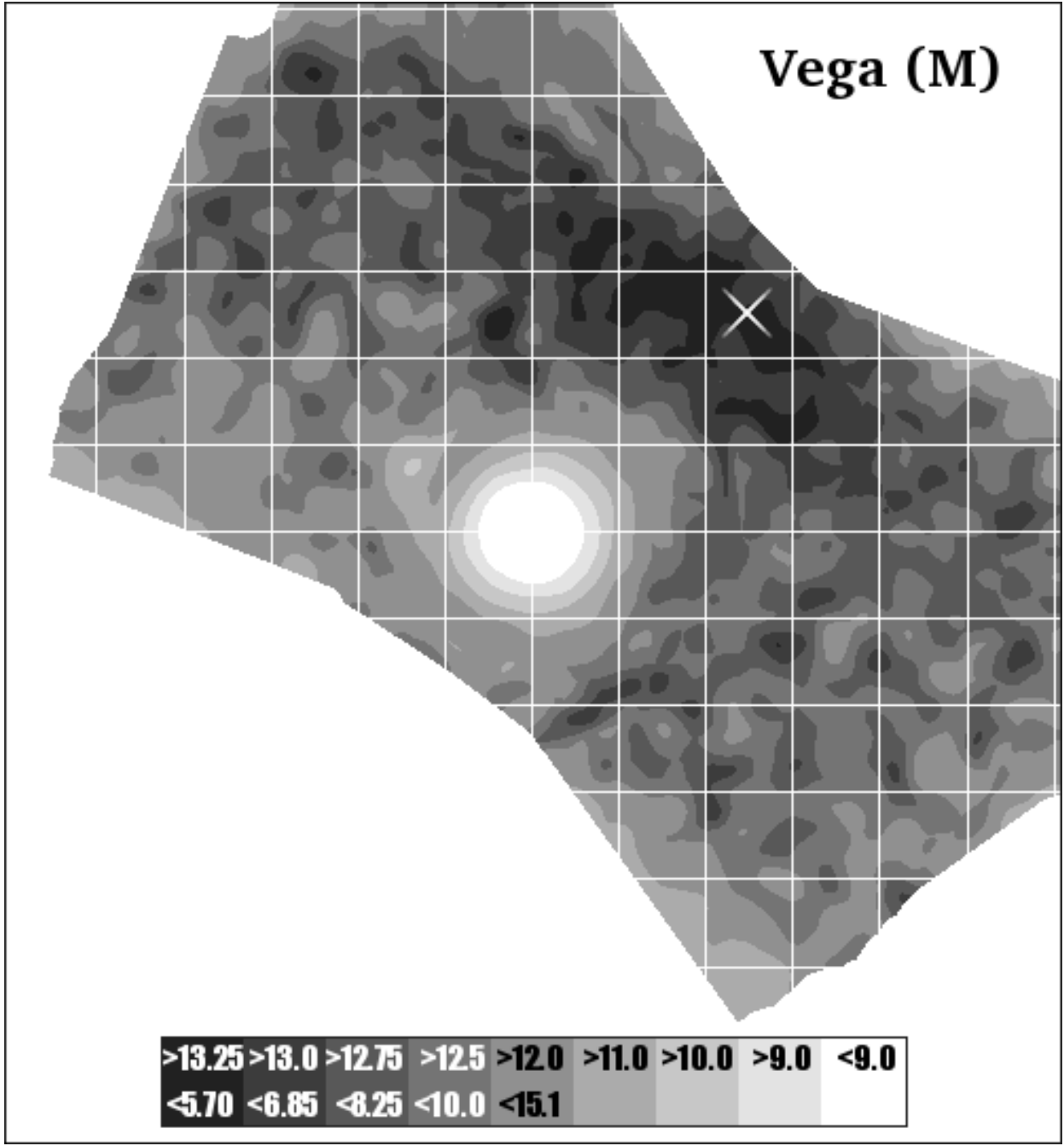} 
\caption[Sensitivity contour map for the Vega $M$ band observations] {
10$\sigma$ sensitivity contour map for our Vega $M$ observations in magnitudes.
The grid squares superposed on the figure for astrometric reference are 2x2 arcsec.
The approximate location of the hypothetical planet from \citet{wilner}
is marked with a white `X'.  The numbers at the top of the colorbar
give the sensitivity of each contour in magnitudes, while the numbers at
the bottom give the equivalent value in MJ, where applicable, 
based on the models of \citet{bur} with the age set to 0.3 Gyr.  }
\label{fig:vegammap}
\end{figure*}

\begin{figure*}
\plotone{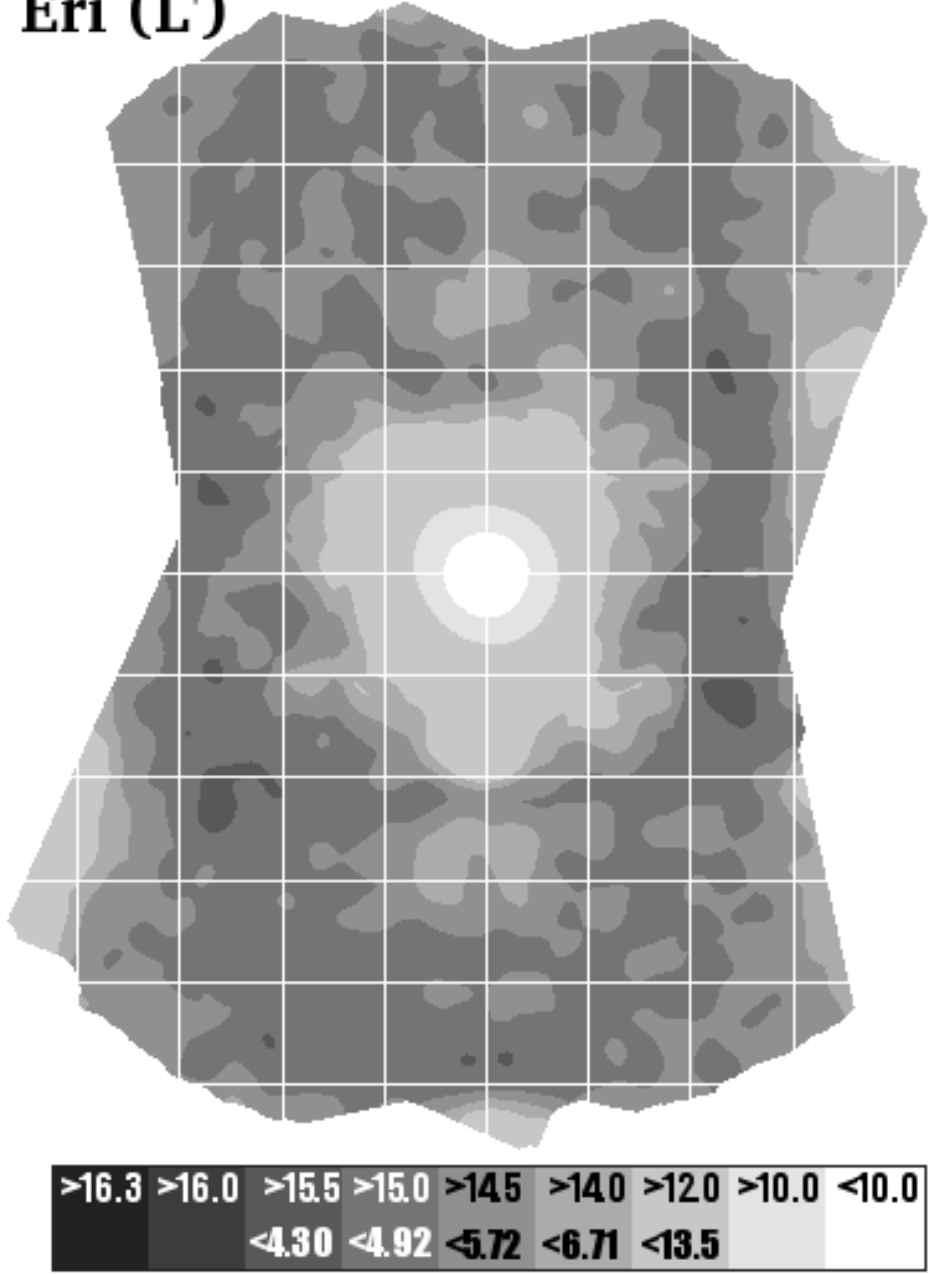} 
\caption[Sensitivity contour map for our $\epsilon$ Eri $L'$ band observations] {
10$\sigma$ sensitivity contour map for our $\epsilon$ Eri $L'$ observations in 
magnitudes.The best areas in this image give sensitivity
to objects fainter than $L'$ = 15.5.
The grid squares superposed on the figure for astrometric reference are 2x2 arcsec. The numbers at the top of the colorbar
give the sensitivity of each contour in magnitudes, while the numbers at
the bottom give the equivalent value in MJ, where applicable,
based on the models of \citet{bur} with the age set to 0.56 Gyr.}
\label{fig:erilmap}
\end{figure*}

\begin{figure*}
\plotone{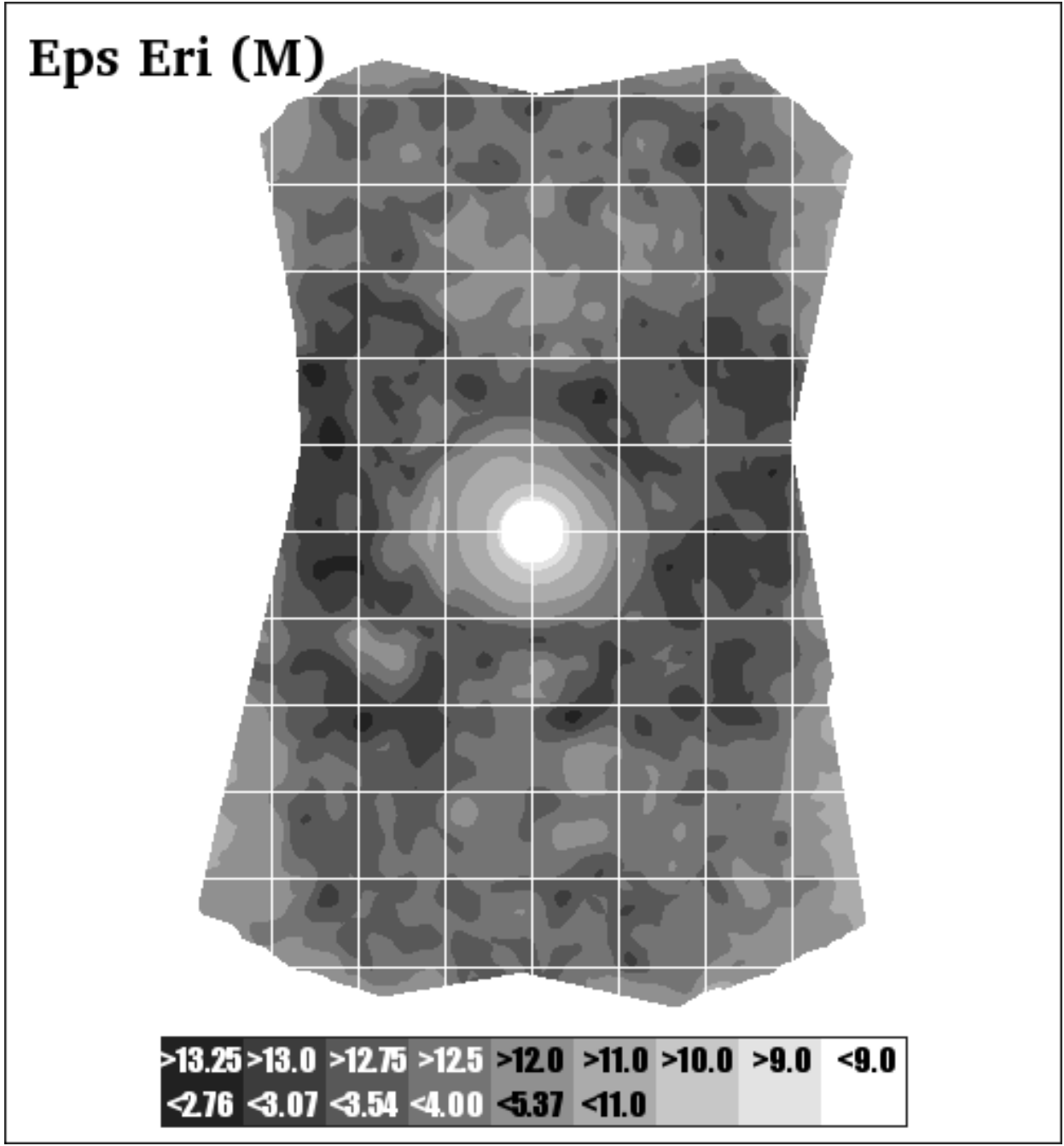} 
\caption[Sensitivity contour map for our $\epsilon$ Eri $M$ band observations] {
10$\sigma$ sensitivity contour map for our $\epsilon$ Eri $M$ observations in 
magnitudes.  The grid squares superposed on the figure for astrometric reference are 2x2 arcsec.  The numbers at the top of the colorbar
give the sensitivity of each contour in magnitudes, while the numbers at
the bottom give the equivalent value in MJ, where applicable, 
based on the models of \citet{bur} with the age set to 0.56 Gyr.}
\label{fig:erimmap}
\end{figure*}

We have converted the master sensitivity maps described above into magnitude
contour images, with 10$\sigma$ sensitivity values shown.  We quote 
sensitivities in apparent magnitudes based on observations of photometric
standard stars (from \citet{leggett}), rather than giving $\Delta$-magnitudes
relative to the primary.  We present our $L'$ and $M$ band Vega results in 
Figures \ref{fig:vegamap}
and \ref{fig:vegammap}, with the approximate position of the hypothetical planet from
\citet{wilner} marked with a white `X'.  Figures \ref{fig:erilmap} and
\ref{fig:erimmap} present the analogous results for $\epsilon$ Eri.
Our Vega $M$ band observation is the deepest ground-based $M$ 
band observation yet published.  

\clearpage

\clearpage

\begin{figure*}
\plotone{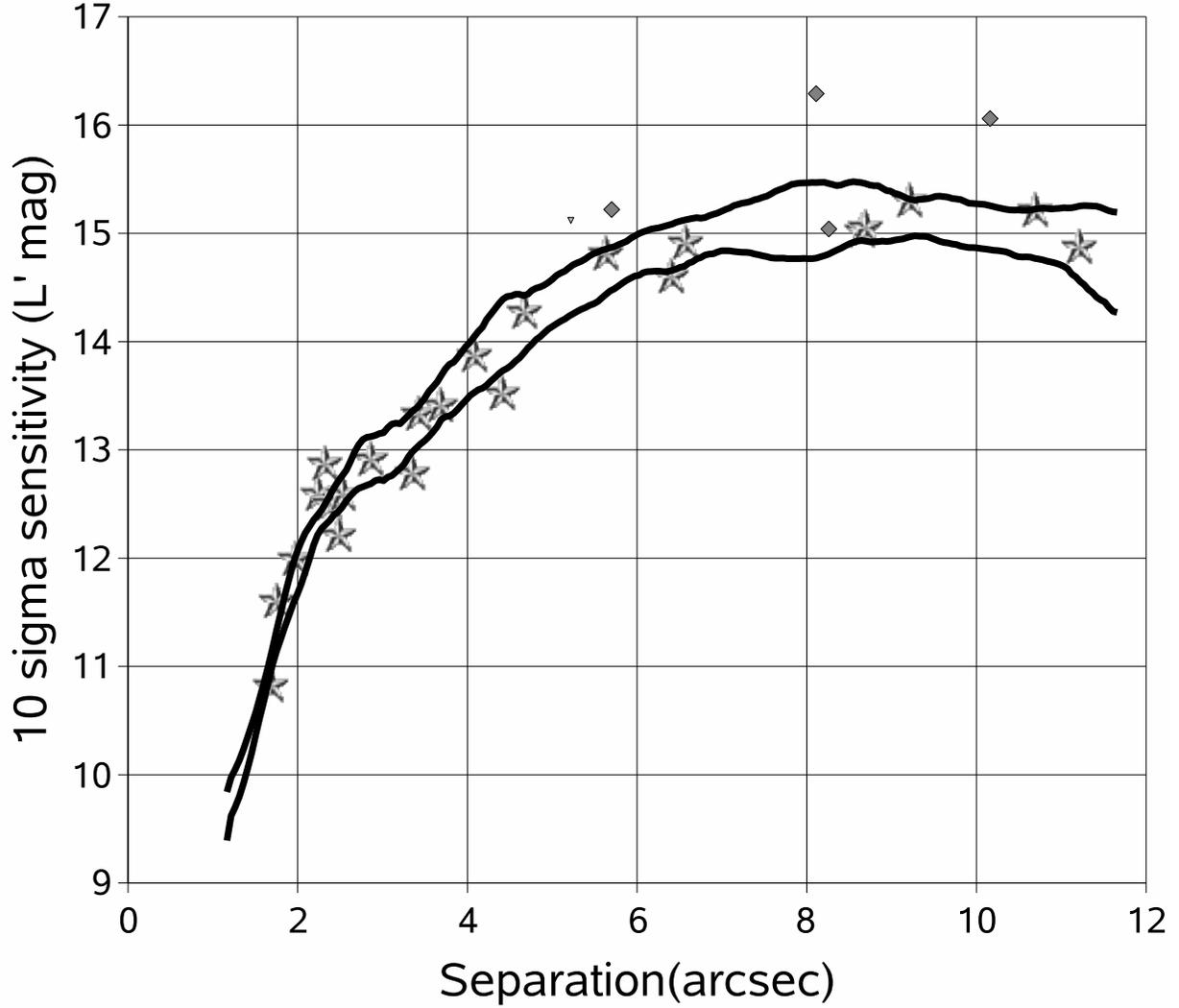} 
\caption[Sensitivity of the Vega $L'$ Band Observations in Magnitudes] {
10$\sigma$ sensitivity of our Vega $L'$ band observations in magnitudes, plotted
against separation in arcseconds.  The 50th and 90th percentile
sensitivity curves are shown, along with fake planets from the
blind sensitivity test.  The star symbols are fake planets that were
confidently detected; the diamonds are those that were suspected
but not confirmed, and the tiny triangle represents the only
fake planet that was not at least suspected.}
\label{fig:vegal_mag}
\end{figure*}

\begin{figure*}
\plotone{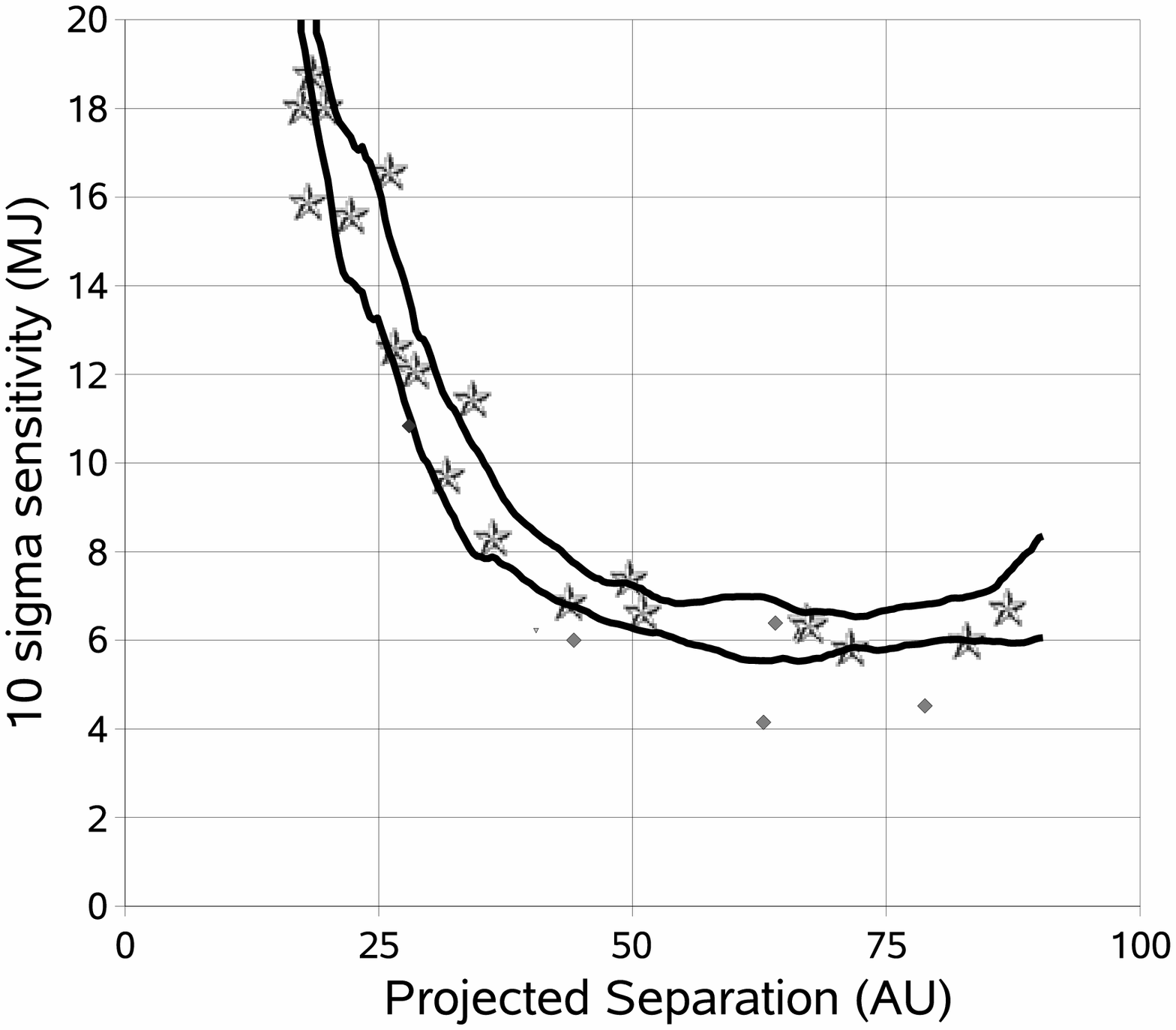} 
\caption[Sensitivity of the Vega $L'$ Band Observations in MJ] {
Sensitivity of our Vega $L'$ band observations in terms of the minimum mass
for a planet detectable at the 10 $\sigma$ level in MJ, plotted 
against projected separation in AU.  
The magnitude-mass conversion was done using the \citet{bur} models for
an age of 0.3 Gyr.
The 50th and 90th percentile sensitivity curves are shown, along with fake planets from the
blind sensitivity test. The star symbols are fake planets that were
confidently detected; the diamonds are those that were suspected
but not confirmed, and the tiny triangle represents the only
fake planet that was not at least suspected.}
\label{fig:vegal_mass}
\end{figure*}

\begin{figure*}
\plotone{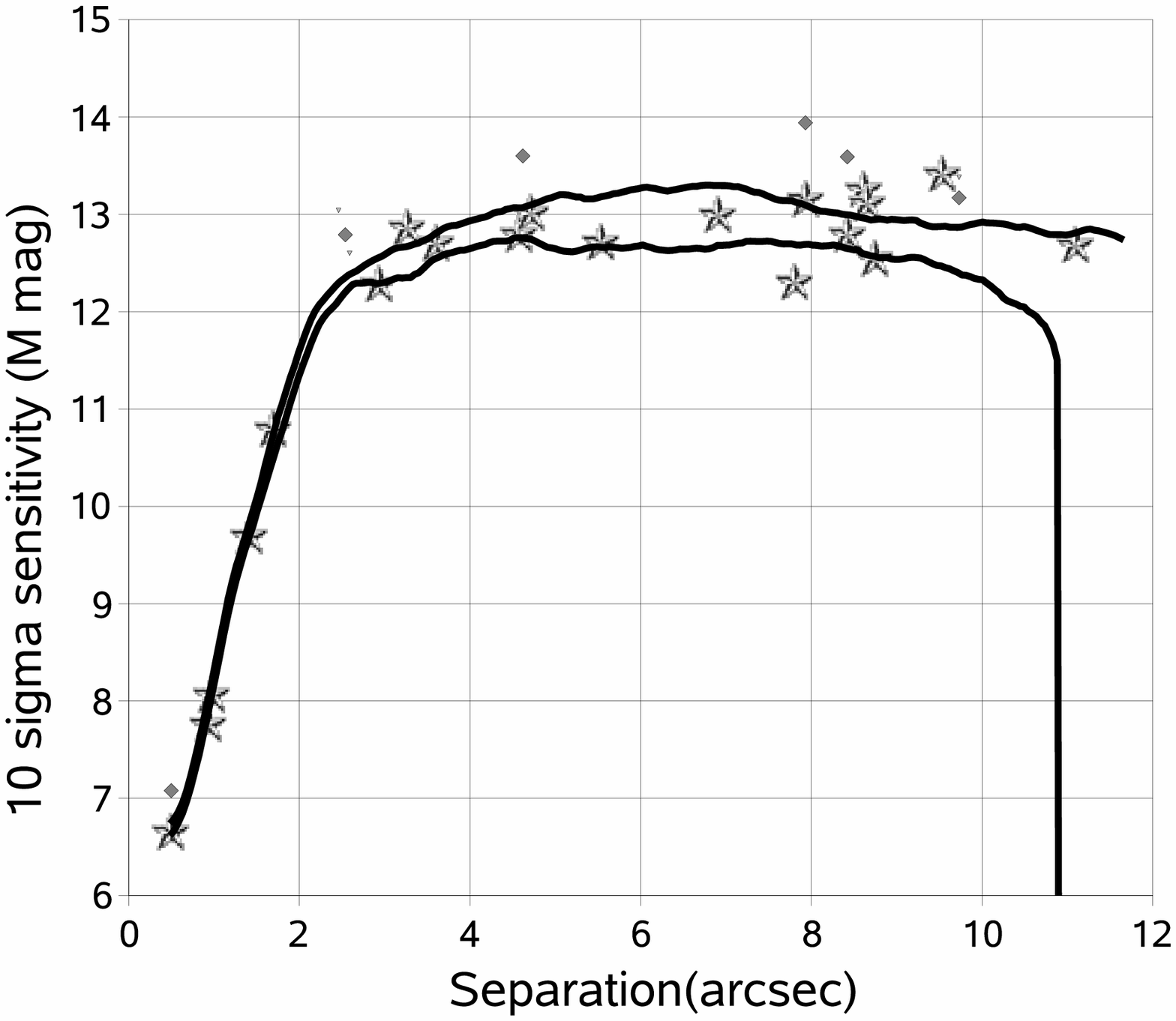} 
\caption[Sensitivity of the Vega $M$ Band Observations in Magnitudes] {
10$\sigma$ sensitivity of our Vega $M$ band observations in magnitudes, plotted
against separation in arcseconds.  The 50th and 90th percentile
sensitivity curves are shown, along with fake planets from the
blind sensitivity test.  The star symbols are fake planets that were
confidently detected; the diamonds are those that were suspected
but not confirmed, and the tiny triangles are those that were
not suspected.}
\label{fig:vegam_mag}
\end{figure*}

\begin{figure*}
\plotone{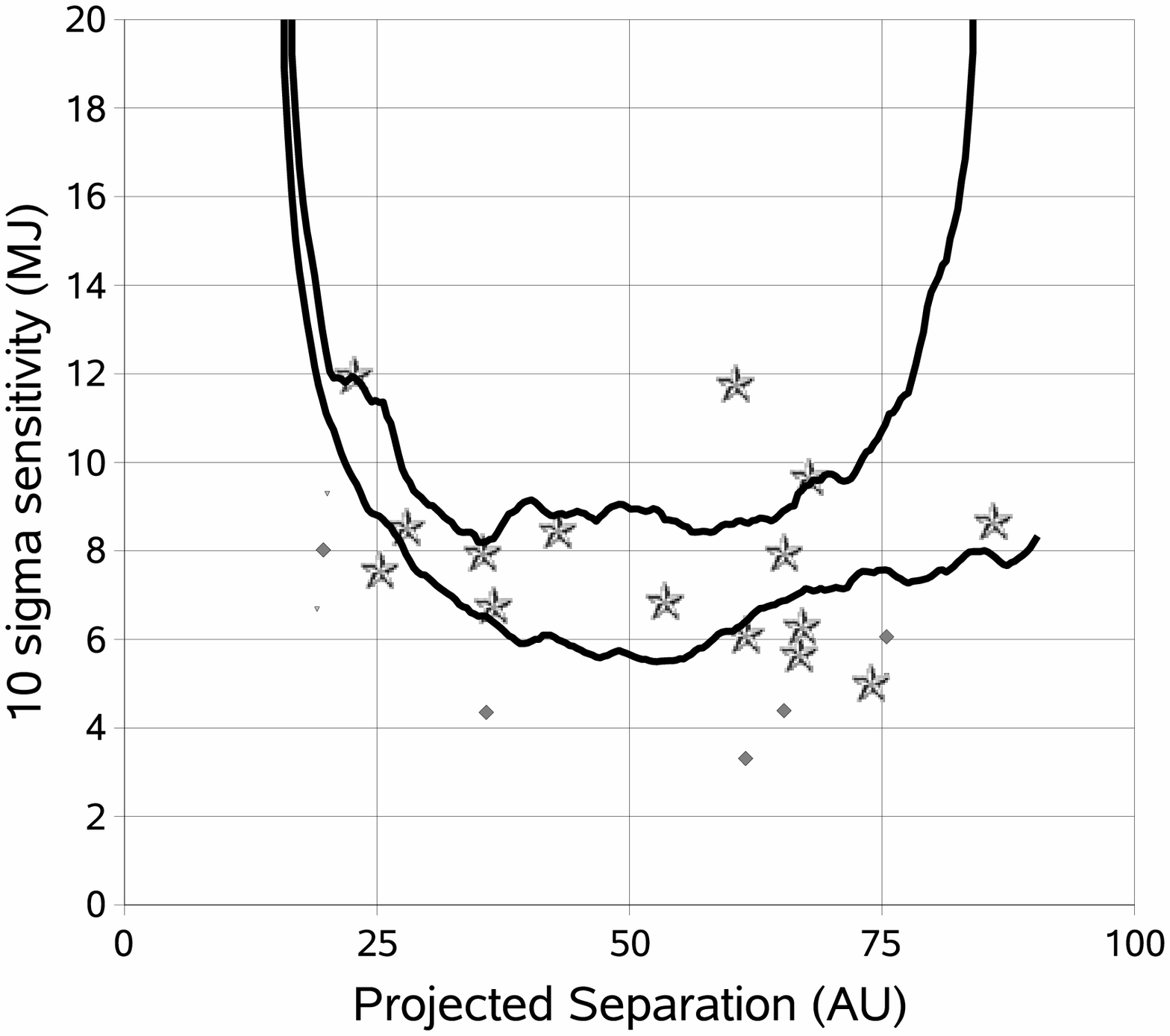} 
\caption[Sensitivity of the Vega $M$ Band Observations in MJ] {
Sensitivity of our Vega $M$ band observations in terms of the minimum mass
for a planet detectable at the 10 $\sigma$ level in MJ, plotted 
against projected separation in AU.  
The magnitude-mass conversion was done using the \citet{bur} models for
an age of 0.3 Gyr.
The 50th and 90th percentile sensitivity curves are shown, along with fake planets from the
blind sensitivity test. The star symbols are fake planets that were
confidently detected; the diamonds are those that were suspected
but not confirmed, and the tiny triangles are those that were
not suspected.}
\label{fig:vegam_mass}
\end{figure*}

\begin{figure*}
\plotone{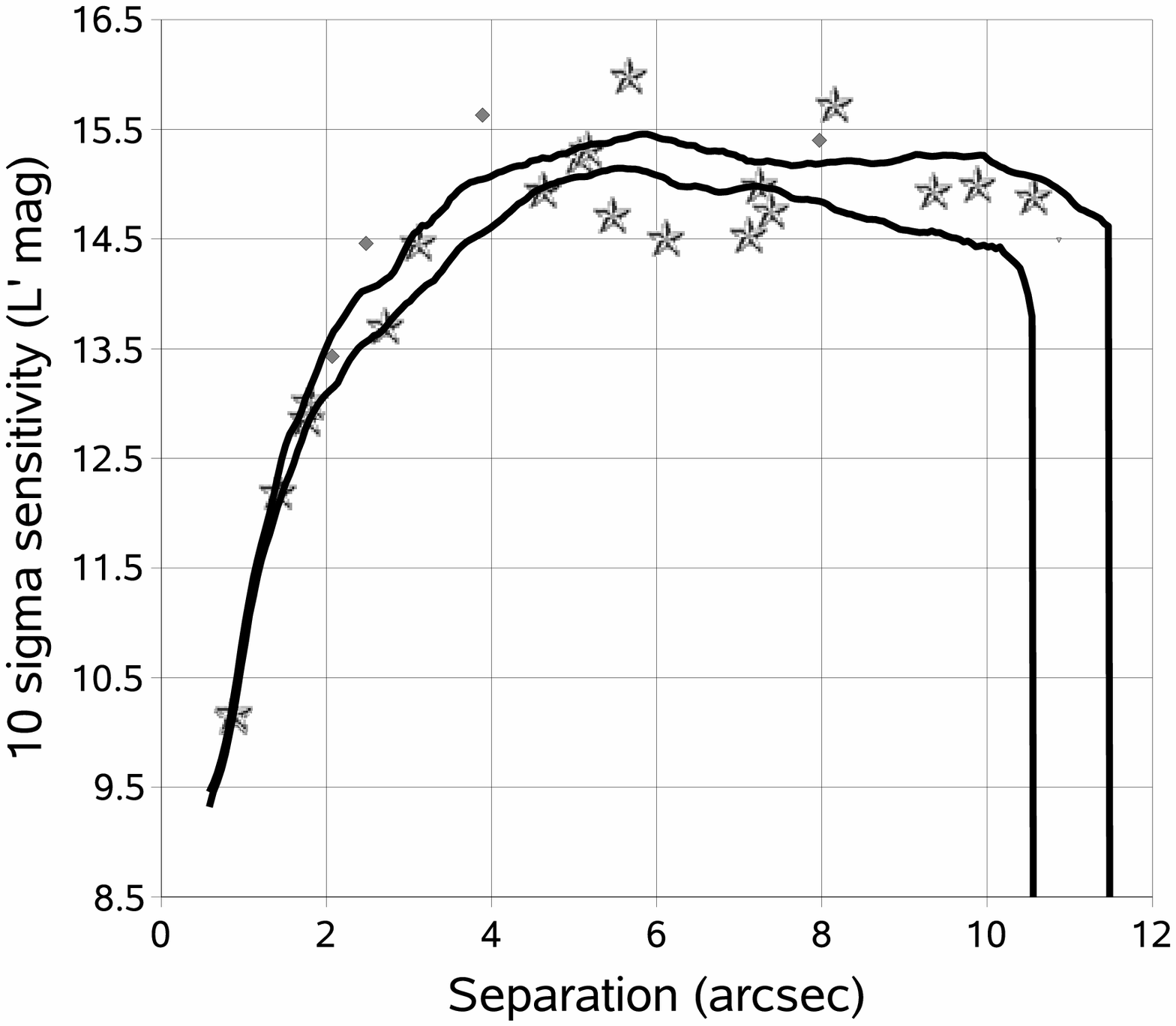} 
\caption[Sensitivity of the $\epsilon$ Eri $L'$ Band Observations in Magnitudes] {
10$\sigma$ sensitivity of our  $\epsilon$ Eri $L'$ band observations in magnitudes, plotted
against separation in arcseconds.  The 50th and 90th percentile
sensitivity curves are shown, along with simulated planets from the
blind sensitivity test. The star symbols are fake planets that were
confidently detected; the diamonds are those that were suspected
but not confirmed, and the tiny triangle represents the only
fake planet that was not at least suspected.}
\label{fig:eril_mag}
\end{figure*}

\begin{figure*}
\plotone{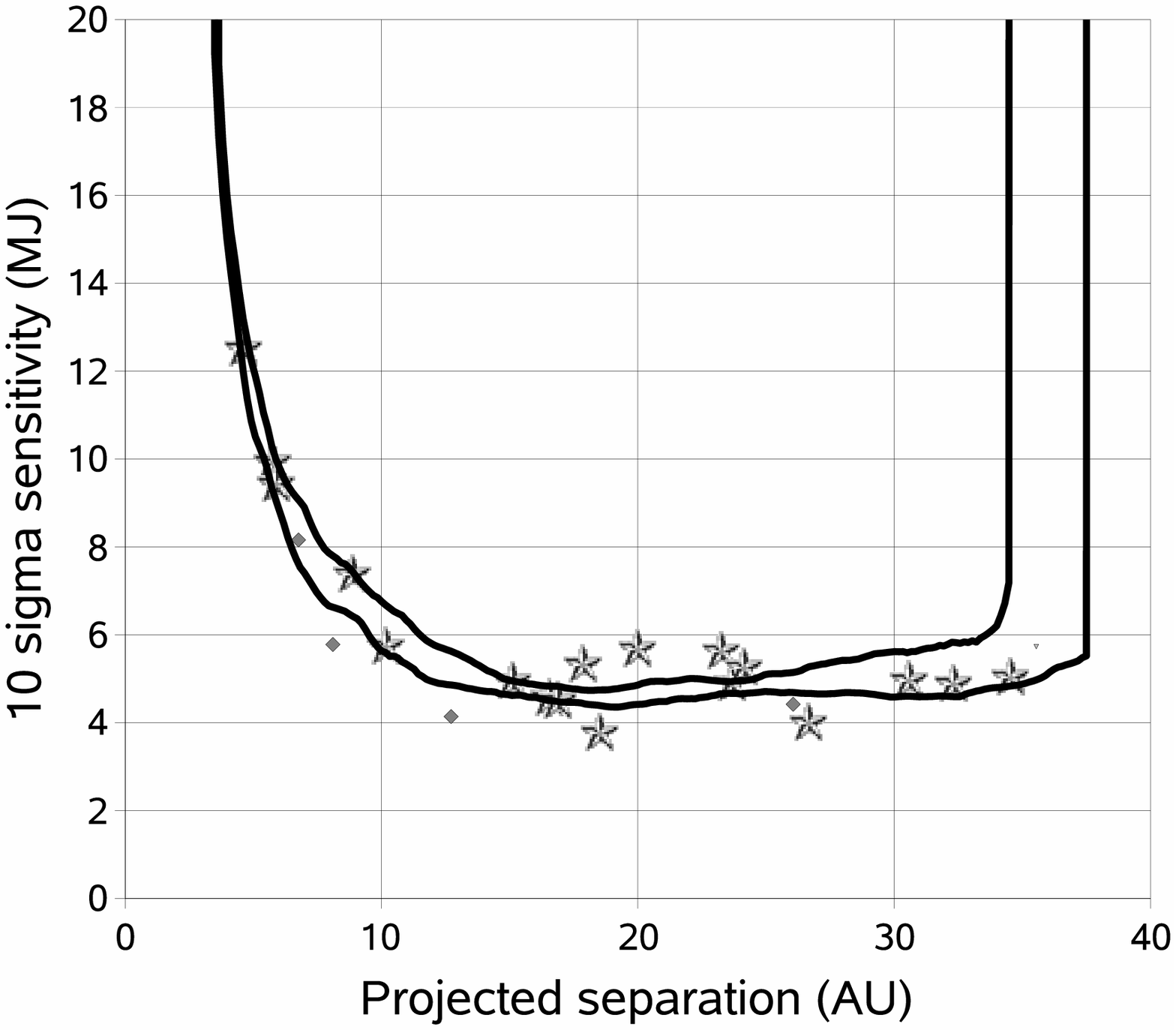} 
\caption[Sensitivity of the $\epsilon$ Eri $L'$ Band Observations in Magnitudes] {
Sensitivity of our $\epsilon$ Eri $L'$ band observations in terms of the minimum mass
for a planet detectable at the 10 $\sigma$ level in MJ, plotted 
against projected separation in AU. The magnitude-mass conversion 
was done using the \citet{bur} models for an age of 0.56 Gyr.  
The 50th and 90th percentile
sensitivity curves are shown, along with fake planets from the blind
sensitivity test. The star symbols are fake planets that were
confidently detected; the diamonds are those that were suspected
but not confirmed, and the tiny triangle represents the only
fake planet that was not at least suspected.}
\label{fig:eril_mass}
\end{figure*}

\begin{figure*}
\plotone{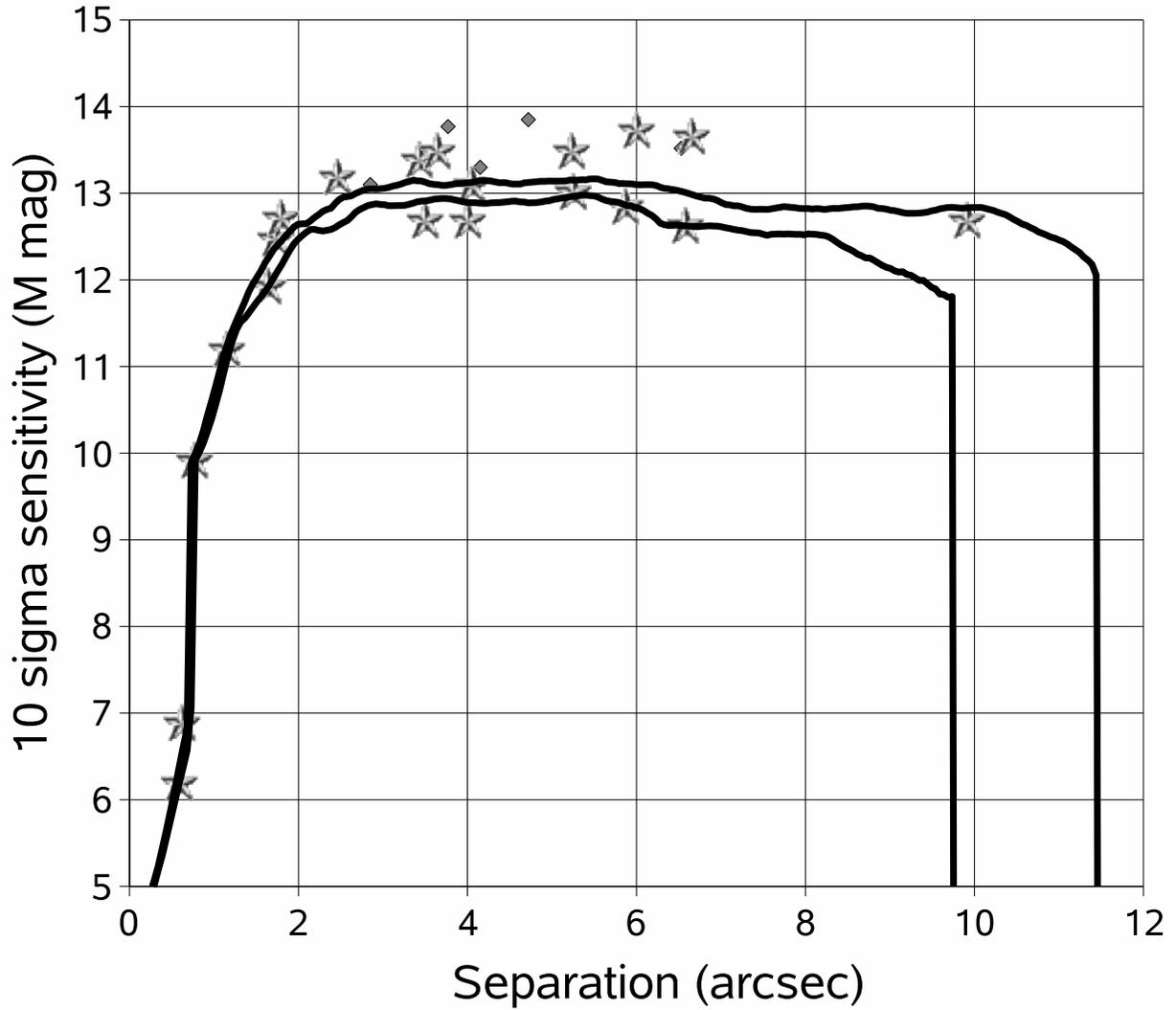} 
\caption[Sensitivity of the $\epsilon$ Eri $M$ band Observations in Magnitudes] {
10$\sigma$ sensitivity of our  $\epsilon$ Eri $M$ band observations in magnitudes, plotted
against separation in arcseconds.  
The 50th and 90th percentile
sensitivity curves are shown, along with simulated planets from the
blind sensitivity test. The star symbols are fake planets that were
confidently detected; the diamonds are those that were suspected
but not confirmed.  In the sensitivity test for this data set all
of the fake planets were at least suspected.}
\label{fig:erim_mag}
\end{figure*}

\begin{figure*}
\plotone{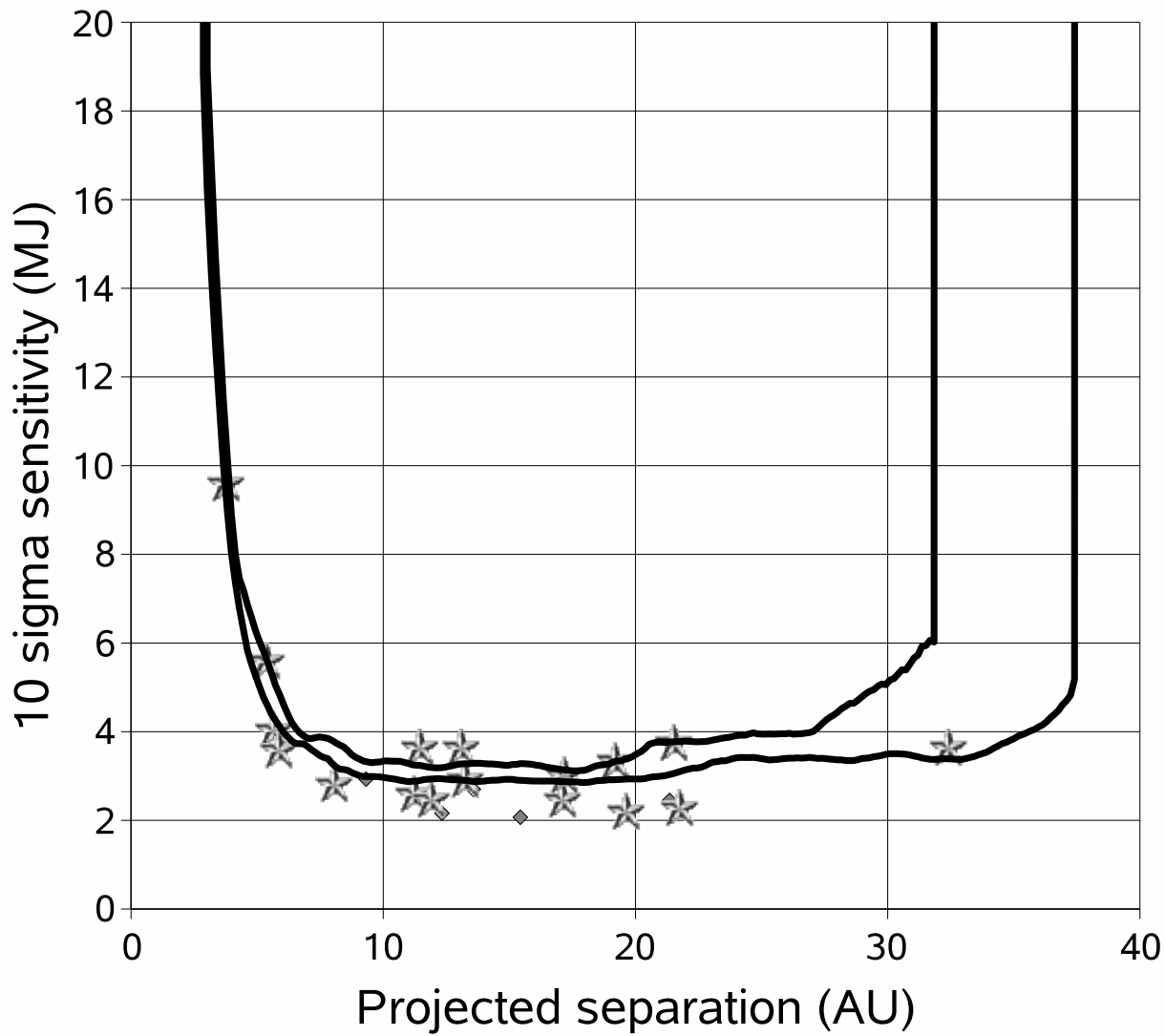} 
\caption[Sensitivity of the $\epsilon$ Eri $M$ band Observations in MJ] {
10$\sigma$ sensitivity of our $\epsilon$ Eri $M$ band  observations in terms
of minimum detectable planet mass in MJ, plotted
against projected separation in AU.  The magnitude-mass conversion 
was done using the \citet{bur} models for an age of 0.56 Gyr.  
The 50th and 90th percentile
sensitivity curves are shown, along with fake planets from the blind
sensitivity test. The star symbols are fake planets that were
confidently detected; the diamonds are those that were suspected
but not confirmed.  In the sensitivity test for this data set all of
the fake planets were at least suspected.}
\label{fig:erim_mass}
\end{figure*}

We have further translated our master sensitivity
map for each data set into 10$\sigma$ sensitivity curves, and plotted them 
in Figures \ref{fig:vegal_mag} through \ref{fig:erim_mass}.  Our sensitivity
varies azimuthally as well as radially due to the negative nod subtraction
images, ghosts, and the different distances to the edge of the valid
data region in different directions.  Therefore we have
computed both the 50th and 90th percentile sensitivities at each radius.
Both are shown, with the 50th percentile, of course, indicating
our median sensitivity and the 90th percentile indicating our sensitivity
in the cleanest 10 \% of the image at a given radius from the star.  
We have also indicated the `Confirmed',
`Noticed', and `Unnoticed' planets from our sensitivity tests with
appropriate symbols.  The sensitivity in these plots increases 
with separation from the star as one would expect, but
then decreases again as the edge of the good data region (ie, useful field on
the master stacked images) is reached.  The noise goes up at
the edge of the useful field because, due to the shifts and rotations
required to register the images, the coverage (number of images
supplying data to a given pixel) goes down near the edge of the field.

Figures \ref{fig:vegal_mag} and \ref{fig:vegal_mass} show the sensitivity we 
obtained in our $L'$ observations of Vega, first in `observational' units
of sensitivity in magnitudes vs. separation in arcsec, and then in `physical'
units of MJ (based on the \citet{bur} models, and adopting an 0.3 Gyr age
for Vega \citep{agevega}) vs. projected separation in AU.
Vega has approximately magnitude 0.0 at every band, so the magnitudes in
Figure \ref{fig:vegal_mag} correspond approximately to $\Delta$-magnitude values.  
Figures \ref{fig:vegam_mag} and \ref{fig:vegam_mass} show the sensitivity
obtained in our $M$ band observations of Vega, following exactly the same
conventions as the $L'$ figures that precede them.

Comparison of \ref{fig:vegal_mass} and \ref{fig:vegam_mass} shows that the
$L'$ and $M$ band results provided similar sensitivity to planets around
Vega.  The $M$ band results are slightly better, especially at
smaller separations.  This is not surprising, because the predicted
planet/star flux ratio is even more favorable at $M$ band than at $L'$.  Also,
MMTAO, like all AO systems, delivers better Strehl ratios at longer
wavelengths, so the PSF subtraction is more effective at $M$ band than
at $L'$.

Figures \ref{fig:eril_mag} through \ref{fig:erim_mass} show 
the sensitivity of our $L'$ and $M$ band observations of $\epsilon$ Eri,
following the same conventions as the Vega figures that precede them.
For $\epsilon$ Eri we have adopted an age of 0.56 Gyr \citep{fischer}.
Note that the magnitudes in Figures \ref{fig:eril_mag} and \ref{fig:erim_mag}
may be converted to $\Delta$-magnitudes by subtracting the $L'$ magnitude 
of $\epsilon$ Eri, which is about 1.72 (the $L'$ - $M$ color of the star
is near zero).

Comparing Figure \ref{fig:eril_mass} with
\ref{fig:erim_mass} shows that for $\epsilon$ Eri the advantage of the
$M$ band over $L'$ is considerably more than for Vega.  The fundamental reason
for this is that $\epsilon$ Eri is closer to us than Vega.  This is
an important point we will refer back to later: the smaller the
distance to a star system, the more favorably long wavelength planet
search observations of the system will compare to short wavelength ones.
There are several logical links in the explaination of this observational fact.
First, intrinsically low-luminosity planets can be detected only
in the nearest systems.  Second, low-luminosity planets have low
$\mathrm{T_{eff}}$.  Third, low $\mathrm{T_{eff}}$ planets have
red $L' - M$ colors. Therefore, the faintest detectable planets
will be more red in nearby systems than in distant ones, and it
follows that longer
wavelength observations (i.e., $M$ band) will perform best relative to shorter
wavelength ones (i.e., $L'$) on the very nearest stars.  This conclusion is 
most obvious when one considers background-limited regions 
of images at large separations from the star, but it applies 
in the contrast limited regime as well.

\clearpage

\clearpage

\section{Vega: Comparison with Other Studies, and Upper Limits for Hypothetical Planets} \label{vega}

\subsection{Comparing Our Sensitivity with Other Studies}
We have not attempted to compare our Vega results with an exhaustive list
of all previous attempts to image planets or other faint objects around
Vega.  Instead, we have chosen two of the best previous results.
First, the $H$ band imaging results of \citet{yoichi}, and second,
the narrow band, $H$-regime images of \citet{marois}.  The latter
presents the most sensitive images yet published
for substellar companions at 3-10 arcsecond separations from Vega.

Before comparing our sensitivities with these other observations a
brief discussion about the different sensitivity estimation techniques
used by the respective observers is in order.  As described above, in this
work we have used an estimator able to account for correlated noise, 
we have performed blind tests of our sensitivity estimator, and we 
have quoted 10$\sigma$ limits.

\citet{yoichi} did not calculate sensitivity limits in terms of $\sigma$.
Instead, they calculated their sensitivities by performing numerous
tests in which they placed 4 planets into their data at a fixed separation
and $\Delta$-magnitude with respect to the primary.  These tests differ
from our own blind sensitivity tests in that the locations of the
\citet{yoichi} fake planets were known, and fixed from one test
to the next.  \citet{yoichi} set their sensitivity at each 
separation to the faintest $\Delta$-magnitude at which at least 
3 of the 4 planets were recovered by their automatic
detection algorithm.  Therefore the \citet{yoichi} sensitivities 
correspond to planet brightness
values at which they had at least 75\% completeness, with an unknown
false-positive rate.  Although it appears the completeness level corresponding
to the \citet{yoichi} sensitivities corresponds better to our 7$\sigma$ level,
we have conservatively chosen to compare the \citet{yoichi} sensitivity 
values to our own 10$\sigma$ results without alteration.

\citet{marois} do not explain how their quoted 5$\sigma$ sensitivity
limits are obtained.  We assume, however, that they used the same method
as \citet{GDPS}, another planet imaging survey by a very similar set
of authors, presenting observations made with the same telescope, instrument,
and observing and analysis stategies. \citet{GDPS} set $\sigma$ 
limits using a sensitivity estimator carefully designed to
account for correlated noise.  They also carefully account for
processing losses, but they do not present blind
sensitivity tests.  Assuming that \citet{marois} used the same
good estimator and careful correction of processing losses, we 
conservatively choose to consider their quoted 5$\sigma$ limits to 
be comparable with our 7$\sigma$ limits.  Based on this assumption we
transform them to 10$\sigma$ limits for comparison with out own.  
We also adjust their limits
by a factor of 2 (0.753 mag) in the direction of greater sensitivity,
to scale from the planet-optimized narrowband filter they used to
the broadband $H$ filter.  (\citet{GDPS} estimate this correction
at a factor between 1.5 and 2.5; we have used the mean value of 2.0.)

\begin{figure*}
\plotone{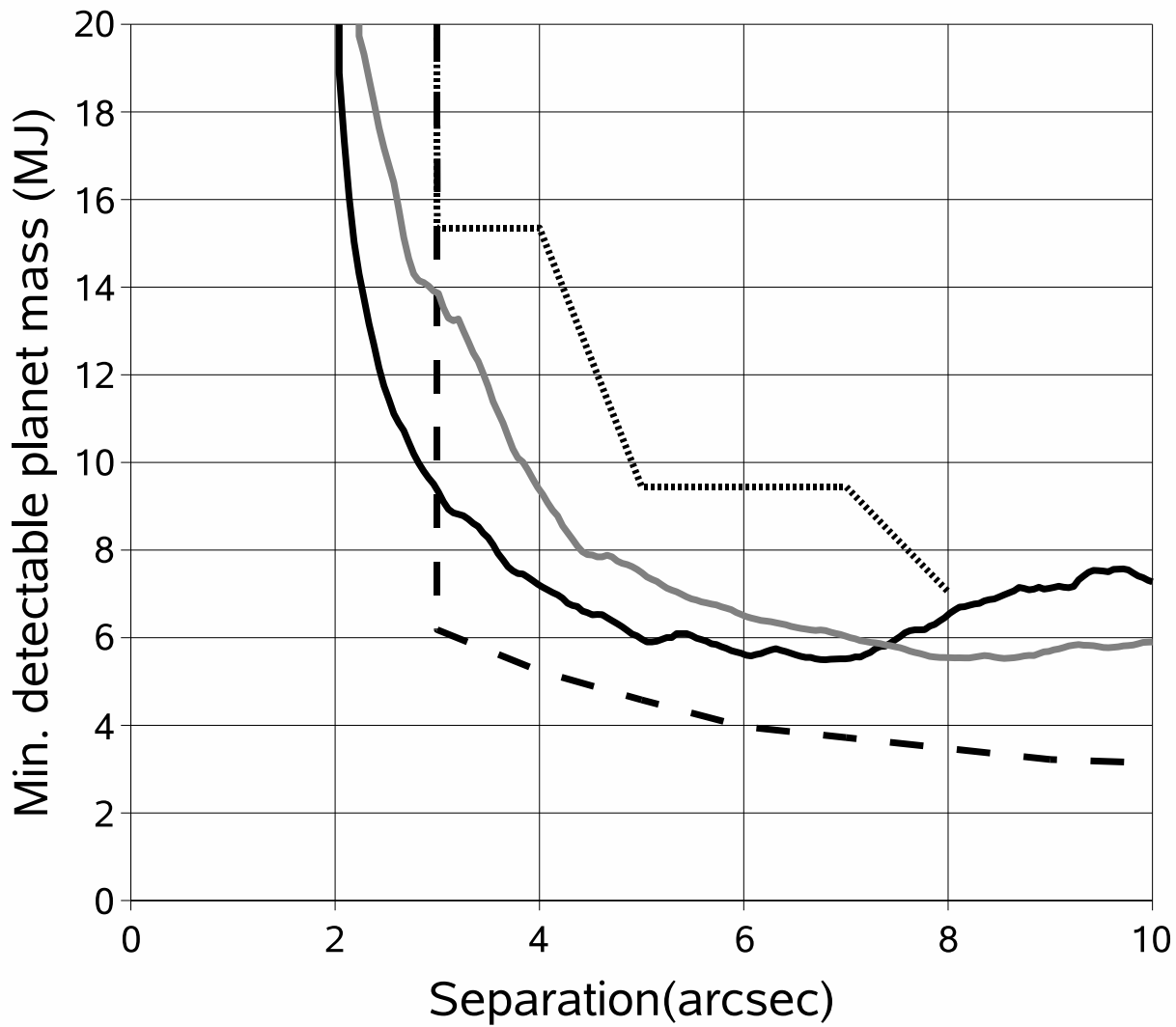} 
\caption[Comparison of sensitivities obtained around Vega with different techniques (using the \citet{bur} models)]{
Comparison of the sensitivities obtained around Vega with different 
techniques.  Magnitude
sensitivities have been converted to planet mass limits in MJ using 
the theoretical
models of \citet{bur} for an age of 0.3 Gyr.  The dashed line is 
the narrowband $H$-regime
result from \citet{marois}; the dotted line is the $H$ band result
from \citet{yoichi}, the gray continuous line is our 90th percentile
$L'$ result, and the black continuous line is our 
90th percentile $M$ band result.}
\label{fig:vhbur}
\end{figure*}

Figure \ref{fig:vhbur} shows the sensitivities of our Vega $L'$ and $M$
band observations compared to those of \citet{yoichi} and \citet{marois}.
The magnitude limits, adjusted as described above, have been
converted to planet masses using the theoretical planet models of \citet{bur},
adopting the \citet{agevega} age of 0.3 Gyr.
We plot our 90th percentile 10$\sigma$ sensitivity values because the 90th
percentile curves are smoother and easier to interpret, and because 
sensitivity at least this good can be obtained at a position angle of 
choice by a well-tuned observing strategy.  Although our observations are
more sensitive to planetary-mass objects
around Vega than the observations of \citet{yoichi}, the carefully processed
narrowband observations of \citet{marois} are more sensitive than ours by
1.5-3 MJ at all separations beyond 3 arcseconds, which was their approximate saturation
radius.  Inward of 3 arcsec our images are sensitive mainly to brown dwarfs
and and the most massive planets, while the other plotted observations
are saturated or very insensitive.  However, in
the regime of higher masses and smaller separations than 
covered by our figure, we note that LYOT project
$H$ band observations of Vega \citep{hinkley} 
obtain sensitivity to massive brown dwarfs inward to about 
0.7 arcsec.  Their observations appear to be sensitive 
to lower mass brown dwarfs than ours inside of 1.5 arcsec, 
while ours are more sensitive at 2 arcsec and farther out.

\clearpage
\clearpage

It is interesting to note that the Figure \ref{fig:vhbur}
would look very different if we plotted $\Delta$-mag rather than
minimum detectable planet mass.  The sensitivity of the $H$ regime 
results of \citet{marois} would surpass the sensitivity of our
observations by a far greater
margin in $\Delta$-mag terms.  At the $L'$ and $M$ bands the sky
background is far brighter than in the $H$ regime.  Also, diffraction
limited resolution is several times lower, and the Airy pattern is 
correspondingly larger in angular terms.  The result is that
despite the cleaner, higher Strehl images offered by AO systems at
longer wavelengths, the $\Delta$-mag vs. angular separation curves at $L'$
and $M$ band are typically considerably less good than those in the $H$
regime.  Because the planet/star flux ratios are so much better
in the $L'$ and $M$ bands, however, when we convert from 
$\Delta$-magnitudes to planet masses the sensitivity gap closes 
considerably, and in fact (as will be seen below in the case of
$\epsilon$ Eri) the longer wavelengths may turn out 
to be more sensitive.

In terms of planet mass the \citet{marois} $H$-band regime observations 
were more sensitive than our $L'$ and $M$ band results beyond 3 arcseconds, 
but not by a huge margin.  Theoretical
planet models are still somewhat uncertain because of the dearth of
observational constraints.  $L'$ and $M$ band observations of bright stars 
such as Vega make sense to diversify the investment of planet-imaging
effort and hedge the overall results against the possibility that
unexpected atmospheric chemistry, clouds, or evolutionary effects
(see for example \citet{faintJup}) cause planets to appear fainter in
$H$ band than current models predict.  It is also possible that planets could be
fainter than predicted at the longer wavelengths, specifically $M$
\citep{L07}.  However, the supression of $M$ band flux observed by
\citet{L07} applied only to objects with $\mathrm{T_{eff}}$ from 700-1300 K.
The situation for objects cooler than 700 K is unknown.
According to the \citet{bur} models, our Vega $M$ band observations 
were sensitive to planets with $\mathrm{T_{eff}}$ below 400 K.  Such objects may be too
cold to have the enhanced concentrations of $\mathrm{CO}$ to which \citet{L07}
attributed the $M$ band flux supression (see \citet{NCE}).

Because they offer better
flux ratios relative to the primary star than shorter wavelengths, 
the $L'$ and $M$ bands we have used are
optimal for detecting massive planets and low mass brown dwarfs
at small separations from Vega and other very bright stars.

\subsection{Upper Limits at the Locations of Hypothetical Planets}
\citet{wilner} presents high-resolution submillimeter observations of 
Vega which show two bright clumps arranged asymmetrically 
relative to the star.  He states that is very unlikely the clumps 
could be background galaxies, and is essentially certain that they 
are concentrations of dust in the Vega system.  Further, the dust 
could represent the remains of two different planetesimal collisions in the
system, but the collisions would have to have happened fairly recently or
the dust would have dispersed.  \citet{wilner} therefore concludes the
most reasonable assumption is that the clumps are dust concentrations
resulting from resonant interactions between the dust and a massive planet.
He shows that the observations could be explained by a 3 MJ planet in
a large, eccentric orbit, which would currently be near apastron and located
about 7.1 arcsec NW of the star (though the submillimeter observations 
were carried out
a few years before our imaging, a planet near apastron in such a large orbit
would not move appreciably over that interval).

We chose the target position and nod direction for our Vega observations
to obtain good sensitivity at the location of this hypothetical
planet.  The planet's location is marked on our sensitivity contour
plots (Figures \ref{fig:vegamap} and \ref{fig:vegammap}).  We do not
detect the planet, so our observations place upper limits on its
mass.

At an approximate separation of 7.1 arcsec, PA 315 degrees (due NW), our $L'$
images of Vega give a 10$\sigma$ sensitivity of $L'$ = 15.21,
or a 7$\sigma$ sensitivity of $L'$ = 15.60.  Translating these
magnitudes to masses using the \citet{bur} models for an age
of 0.3 Gyr, and using the results
of our blind sensitivity tests, we can rule out a planet at this location
with a mass above 6.02 MJ with near 100\% confidence, and one more massive than
 4.30 MJ with 77\% confidence.  If the images were very clean at this
location, showing no suspected sources, we could set stronger limits.
However, there was a suspected source within about 0.4 arcsec of this
location.  Careful records of the manual examination of the images make
it clear that the suspected source can be identified as spurious
with high confidence, and should by no means be considered a candidate 
detection of the \citet{wilner} planet.  Its appearence simply means
the images are not very clean at this location, and the stronger
limits possible in regions without suspected sources do not apply.

At the same location on our $M$ band images, we
obtained 10, 7, and 5$\sigma$ limits of $M$ = 13.39, $M$ = 13.78,
and $M$ = 14.14, respectively.  Using the \citet{bur} models for
an age of 0.3 Gyr, these 
magnitude limits correspond to planets of 5.14 MJ, 3.76 MJ, and 2.86 MJ,
respectively.  Records from our automatic and manual examination 
of the images show no suspected source within
1.5 arcseconds of this location.  Since in the sensitivity tests
97\% of 7$\sigma$ planets and 85\% of 5$\sigma$ planets were at
least suspected, we can rule out a planet above 3.76 MJ at this
location at the 97\% confidence level, and one above 2.86 MJ at
the 85\% confidence level.  The excellent sensitivity obtained at
this location is due in part to the fact that our
observing strategy was optimized to give good sensitivity near
the position of the \citet{wilner} hypothetical planet.


We can set limits on the hypothetical planet of \citet{wilner}
close to, or perhaps even below, the proposed mass of 3 MJ.  It would 
appear from Figure \ref{fig:vhbur} that \citet{marois} set
similar or slightly lower limits, though the exact sensitivity
of their observations at the position angle of the 
\citet{wilner} planet cannot be explicitly
analyzed because they present their sensitivity only in a radially
averaged sense.  Observations at the $H$ and $M$ bands have thus consistently 
set upper limits near the predicted mass of 3 MJ.  A 3 MJ planet at 
the 0.3 Gyr age we have adopted for Vega would have $\mathrm{T_{eff}}$ 
between 300 and 400 K.  No objects in this temperature range have 
yet been observed, so model fluxes are not observationally constrained
at any wavelength.  Where an upper limit from a single band would
be tentative because of the uncertainties of the models, the consistent 
results from a range of wavelengths allow us to conclude that it is
probable no 3 MJ planet exists at this location.

\citet{wilner} makes it clear that other models besides his hypothetical 3 MJ planet
might explain the observed dust distribution, and
that further modeling is needed to see what range of planetary orbits
and masses might be capable of producing the resonant dust concentrations
seen in the submillimeter.  \citet{marsh}, for example, explain the
distribution of dust they observe around Vega at 350-450 $\mu$m wavelengths 
(vs 850 $\mu$m for \citet{wilner})
by a Neptune-mass planet in a 65 AU orbit.  It is not entirely clear whether
their model also explains the \citet{wilner} images; however, \citet{wyatt}
presents a model of a migrating Neptune-mass planet that does match the 850 $\mu$m
images.  In contrast to \citet{wyatt} and \citet{marsh}, \citet{deller} present 
a model that explains
the 850 $\mu$m images by a 3 MJ planet in a considerably larger orbit than
that suggested by \citet{wilner}.  It would have the same current PA
as the \citet{wilner} planet (NW of the star, near PA $315^{\circ}$), but
it would be 12-13 arcsec from Vega as opposed to 7 arcsec.  Our Clio observations
do not obtain good sensitivity at these larger separations, though new, differently
targeted Clio images could.

No current observational technique can image Neptune-mass extrasolar planets in distant
orbits.  The non-detections of our survey and that of \citet{marois} lend
some support to models explaining the Vega dust distribution using such
planets rather than the model of \citet{wilner} in which the planet
has a mass a few times that of Jupiter.  However, we cannot rule out a 3 MJ
planet in the more distant orbit suggested by \citet{deller}, simply
because our observing strategy was not designed to give good sensitivity
at such a large separation.  

Theoretical planet models indicate that observations at $L'$, $M$ band, and 
the narrowband $H$-regime filter of \citet{marois} can
detect planets down to 3 MJ in the Vega system.  Further work
at all three bands would either detect such a planet or rule out the 
existence of one at large separation with very high confidence.
Consistent results at a variety of wavelengths will ensure that conclusions
are less vulnerable to model uncertainties at any particular wavelength.  
More submillimeter work and orbital modeling of the 
Vega system is also desirable, because if models explaining the dust
distribution without a massive planet can be ruled out, deep
targeted AO observations to detect the planet could be strongly
prioritized, and success could be anticipated with confidence.

\section{$\epsilon$ Eri: Comparison with Other Studies, and Upper Limits for Hypothetical Planets} \label{eri}

\subsection{Comparing Our Sensitivity with Other Studies} \label{compsens}
\begin{figure*}
\plotone{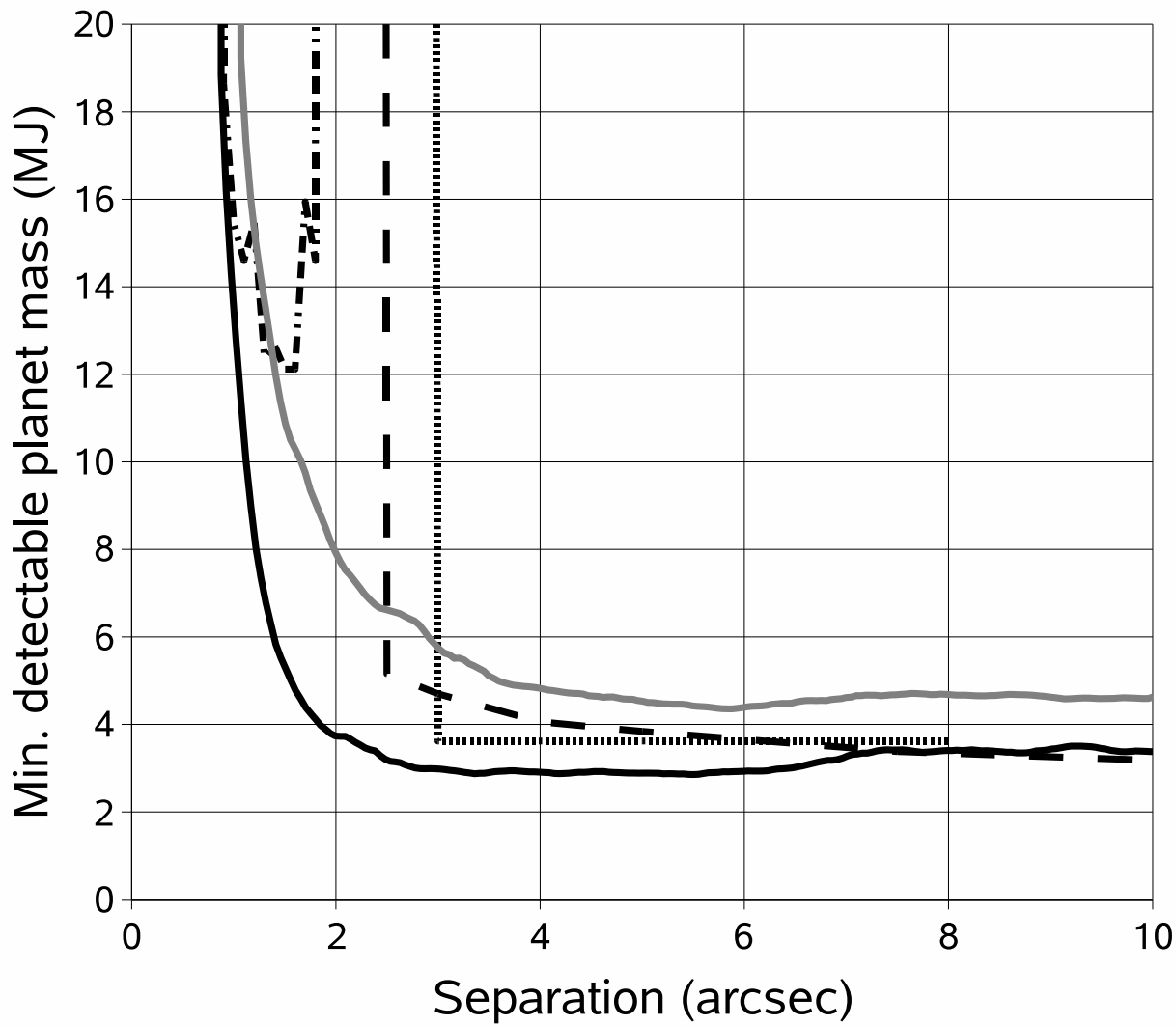} 
\caption[Comparison of sensitivities obtained around $\epsilon$ Eri with different 
techniques (using the \citet{bur} models)]{
Comparison of the sensitivities obtained around $\epsilon$ Eri with different 
techniques.  Magnitude
sensitivities have been converted to planet mass limits in MJ using 
the theoretical
models of \citet{bur} for an age of 0.56 Gyr.  The long-dashed line 
is the narrowband $H$-regime
result from \citet{GDPS}, the dot-dashed line at small separations 
is the SDI result from \citet{biller}, the dotted line is the $H$ band result
from \citet{yoichi}, the gray continuous line is our 90th percentile
$L'$ result, and the black continuous line is our 
90th percentile $M$ band result.}
\label{fig:ehbur}
\end{figure*}

As with Vega we do not attempt to compare our $\epsilon$ Eri
results with an exhaustive list of other studies, but only
with a few that obtained the best sensitivity results.
We have chosen \citet{yoichi}, \citet{biller}, and \citet{GDPS}.
Figure \ref{fig:ehbur} shows the results of the comparison, with
again, the 90th percentile 10$\sigma$ sensitivity curves for our observations
plotted.

Of the other studies, the sensitivity methods of \citet{yoichi} have 
already been discussed in 
Section \ref{vega} above, as have those of \citet{GDPS} because we
assumed \citet{marois} used the same methods for their Vega data.
It only remains to consider the methods of \citet{biller}.
They use a sensitivity estimator which is based on the single-pixel
RMS in 6 pixel (0.05 arcsec, or 1.2 $\lambda / D$) square 
boxes on the images, and they quote 5$\sigma$ limits.  It is not
clear whether they take processing losses into account in their
sensitivity calculation.  In general we expect sensitivity estimators
involving the single-pixel RMS to overestimate the sensitivity, as
they assume independence of noise in adjacent pixels.  This assumption
is always violated in the speckle-dominated regions on AO images (that
is, speckle noise is always spatially correlated, though the extent
of the correlation depends on the details of the raw images
and the type of PSF subtraction used).  

The above would seem to imply that the \citet{biller} 
5$\sigma$ sensitivity results are comparable to our 5$\sigma$
limits, and that we should adjust them by a factor of 2 (0.753 mag) toward
decreased sensitivity in order to compare them properly against
our 10$\sigma$ limits.  This would not include any correction for
the possible overestimation of sensitivity in the presence of correlated
noise.  

However, several characteristics of the \citet{biller} data suggest
their sensitivity should be rated higher than this.  First, they
use a `roll subtraction' technique which effectively
creates both a positive and a negative image of any
real companion, separated by $33^{\circ}$ of rotation about the primary
star, and the presence of both can be used to evaluate
the reality of potential sources.  This doubles the
data and the sensitivity should accordingly go up by
$\sqrt{2}$.  

Second, their simultaneous differential 
imaging (SDI) technique involves two independent spectral differences.
They are not necessarily equally sensitive, but in the best case 
this again doubles the data available for planet detection.
With, potentially, four equal-brightness images of any
real object in their data (two independent spectral difference images
at each of two `roll angles'), the
sensitivity of the \citet{biller} observations should in
principle go up by a factor of as much as 2 (i.e. $\sqrt{4}$)
over their nominal values.  

Finally, Beth Biller has explained 
to us that the \citet{biller} 5$\sigma$ point-source sensitivities 
were calculated by comparing the single-pixel RMS noise
to the brightness of the peak pixel of a PSF.  This method is
conservative for well-sampled data such as that of \citet{biller},
since it does not take into account the fact that bright pixels
surrounding the peak of a PSF allow it to be detected with additional
confidence.  The single-pixel method also does not overestimate 
the sensitivity in the presence of correlated noise (provided the
RMS noise is calculated over a large enough region).

The above might indicate we should compare the \citet{biller}
nominal 5$\sigma$ sensitivities directly to our 10$\sigma$ sensitivities
(since obtaining 4 separate images of any real source could
in principle raise the sensitivity to twice its nominal value).  
However, since the two spectral difference images do not neccesarily
have equal sensitivity, we have scaled the \citet{biller}
nominal 5$\sigma$ limits down in sensitivity by about a factor of 
$\sqrt{2}$ (0.38 mag) to compare them with our 10$\sigma$ 
limits.  This is equivalent to taking into account the \citet{biller}
sensitivity gain only from the fact that an image is obtained
in each of two `roll angles', and not from the additional fact
that at each roll angle two independent spectral difference images
are produced.  The reader should keep this in mind when examining
Figure \ref{fig:ehbur}: we may have underestimated the relative
sensitivity of the \citet{biller} obsevations by a factor of 
around $\sqrt{2}$ (0.38 mag).  This rather small correction would
not affect our conclusions.

As with the \citet{marois} Vega data (and also the \citet{GDPS} $\epsilon$
Eri data), we have adjusted the \citet{biller}
sensitivities toward greater sensitivity to convert 
magnitudes from narrowband filters tuned to a predicted peak 
in giant planet spectra to broadband $H$ magnitudes.  For \citet{biller},
the correction factor we applied was 0.84 magnitudes.  This is an 
approximate value based on the SDI observers' analysis of their own filters.

\clearpage
\clearpage

Figure \ref{fig:ehbur} makes it clear that though the best $H$-regime
results for Vega delivered better sensitivity than our $L'$ and $M$ band
observations, the sensitivity of our $M$ band observation of $\epsilon$ Eri
is better than that of all previous observations out 
to a separation of at least 7 arcseconds from the
star.  Within three arcseconds of the star the sensitivity advantage
of the longer wavelength observation is especially great.  We note
that this applies only to our $M$ band result: the SDI method of 
\citet{biller}, which is designed to give excellent sensitivity 
close to bright stars, does give results comparable those of 
our $L'$ observation.  The good performance of the
$M$ band is due to the fact that the planet/star flux ratio 
is much more favorable at $M$ band than even in the most 
optimized intervals of the $H$ band.

In closing this section on comparitive sensitivities, we note 
that although we have used mainly the theoretical planet 
models of \citet{bur} to calculate sensitivities in this work, those
presented in \citet{bar} are a good complement and comparison 
to the former.  However, the filter set over which \citet{bar} 
integrated their theoretical spectra is slightly different from 
the Clio filter set, over which we integrated the \citet{bur} models.  
We have, however, also done tests in which we performed 
magnitude-mass conversions using the original mass/mag/age 
tables presented in \citet{bar}.  In general, the \citet{bar} models
give us somewhat better sensitivity in $L'$ than those of \citet{bur}, 
with a typical disagreement of 1-2 MJ.  The $M$ band predictions of the
two model sets are very close.  At present we cannot say for sure if
the $L'$ band discrepancy is inherent in the different models or is
a pure artifact of the filter set.  In any case the two model sets
are broadly in agreement, except for very old, cool planets, where the
differences become very large and it appears clear that slightly different
filter sets cannot be the whole explanation (see the discussion of the $H$ band
flux of $\epsilon$ Eri b below).

\subsection{Upper Limits at the Locations of Hypothetical Planets}

$\epsilon$ Eri has the extremely important distinction of being
one of only a few stars around which a single planet has been detected with both RV and
astrometric methods \citep{hatzes, benedict}.  This means that a complete, unique
solution for the size, eccentricity, and orientation of the orbit is possible,
as is a solution for the mass of the planet.  \citet{benedict} present
such an orbit solution, and give the mass of the planet as 1.55 MJ.

At the time of our observations the \citet{benedict} orbit predicts a separation
of about 0.684 arcsec.  Our observations do not set limits in the
planetary mass regime this close to the star.  We note, however, that our observations
were not timed with the idea of obtaining good sensitivity to this
planet.  If we had observed the planet near its apastron, at which
point the separation is about 1.7 arcsec, our $M$ band observations
in particular would have been in the range to set possibly interesting
limits, though still above the \citet{benedict} mass of 1.55 MJ.
The median 10 and 7$\sigma$ sensitivities of our $M$ band observation at
1.7 arcseconds are 5.3 and 4.2 MJ, respectively, and our 5$\sigma$ limit
is 3.9 MJ.  These are good sensitivities at a very small separation
from a bright star, but, of course, the planet would still not have been 
detected, unless it is far more massive than the 
\citet{benedict} orbital solution indicates.

Could any current-technology telescope detect this planet, and if so
what would be the best method?

\citet{janson} applied the same SDI methodology used by \citet{biller} 
to observe $\epsilon$ Eri at several different epochs.  
The data from their
second epoch gave them the best limit on the planet, with a 3$\sigma$ 
sensitivity of $\Delta$-magnitude 13.1 at the expected location
of the planet based on \citet{benedict}.  As discussed above, the
\citet{biller} observations using the SDI method had two independent
roll angles and two independent spectral differences for each observation,
and the sensitivity estimation method they used was conservative.
Assuming that \citet{janson} used the same methodology, we will compare
their 3$\sigma$ limits directly to our 10$\sigma$ limits.  Note that
even considering all the issues mentioned in Section \ref{compsens}
above this results in a conservative estimation of our 
sensitivities relative to those of \citet{janson}.

We can adjust the \citet{janson} 3$\sigma$ 
sensitivity of $\Delta$-magnitude 13.1 by the 0.84 mag
value used before and add the $H$ = 1.88 magnitude of the
star itself to get an equivalent sensitivity of $H$ = 15.8;
the equivalent masses are 9.6 and 9.1 MJ according to the \citet{bur}
and \citet{bar} models, respectively, with the age set to 0.56 Gyr
in both cases.

However, \citet{janson} mention that the correction from the
narrowband SDI filters to $H$ band would actually be much greater 
than 0.84 magnitudes for a very cool object such as $\epsilon$ Eri b.  According to
their Figure 5, the correction is about 2.2 magnitudes for
the appropriate filter difference at our adopted age of 0.56 Gyr
for $\epsilon$ Eri.  This different correction does not
change the upper limit of 9.6 MJ quoted above, because the
larger correction applies only to a planet with the 1.55 MJ
mass determined by \citet{benedict}, which would have been
far too faint for \citet{janson} to detect.  However, the 2.2 magnitude
correction is appropriate for estimating by what factor 
the \citet{janson} observations missed the planet -- that is, how
much their sensitivity would have to be increased in order to
detect it.

The sensitivity of the \citet{janson} observations 
in their narrowband filter was about
13.1 + 1.88 = 14.98 mag, assuming that the magnitude of $\epsilon$ Eri A
is the same in the narrowband filter as in broadband $H$.  According
to the models of \citet{bur}, an 0.56 Gyr-old planet of mass 1.55 MJ
located 3.27 pc away has an $H$ band magnitude of about 28.5.  We subtract
the 2.2 magnitude correction to obtain a narrowband magnitude of 26.3, 
and difference the result with the \citet{janson} sensitivity of 
14.98 mag.  The conclusion is that the \citet{janson} sensitivity 
was insufficient to detect the planet by 11.3 magnitudes (a factor of 34,000) under the
\citet{bur} models.  As noted in Section \ref{compsens} above, the \citet{bar} models
disagree with the \citet{bur} ones on the brightness of $\epsilon$ Eri b:
the former indicate that the \citet{janson} miss factor is only
about 1000, rather than 34,000.  The reason for this discrepancy
is not clear.  It does not affect our conclusions about the best
band at which to search for $\epsilon$ Eri b.  The large
discrepancy for two model sets that are in close agreement for warmer
objects does suggest that theoretical $H$ band magnitudes for
objects with temperatures as low as $\epsilon$ Eri b have a large uncertainty,
and therefore any constraints based on them will be tentative.  
The $M$ band brightnesses predicted by the two
model sets for $\epsilon$ Eri b are discrepant by a much smaller
factor, about 1.7 rather than 34 (see below).

The miss factors calculated above indicate the SDI sensitivity
would have to be increased at least a thousandfold to
detect the planet.  Assuming we
had observed the planet at apastron, by what factor would we
have failed to detect it?  We will consider only our $M$ band
results, as they are more sensitive than our $L'$ observations
to low mass planets close to the star.  
Our median 10$\sigma$ sensitivity at the apastron separation
of 1.7 arcsec was $M$ = 12.02.  The models of \citet{bur} give
the brightness of the planet as $M$ = 14.7.  This means we 
would have come short of a 10$\sigma$ detection by
2.68 mag, or a factor of 11.8, according to the \citet{bur} models
(the \citet{bar} models give a higher but not enormously discrepant
miss factor of 20.5; as in the paper up to this point we focus on
the \citet{bur} models in the discussion that follows).
Our blind sensitivity tests indicate about 40\% completeness
at 5$\sigma$, with 85\% of sources at least noticed.  
Thus if we could increase our sensitivity
by only a factor of 5.9 (that is, 11.8 divided by 2 to change
from 10 to 5$\sigma$), we would have some chance of confidently
detecting the planet, with a greater likelihood of at least noticing it.

These lower miss factors suggest that $\epsilon$ Eri b
might actually be detectable near apastron with ground based 
$M$ band imaging.  It is almost certain that $\epsilon$ Eri b
is at too low a $\mathrm{T_{eff}}$ for its $M$ band flux to
be dimmed by the above-LTE $\mathrm{CO}$ concentrations suggested
by \citet{L07} and \citet{reid} to account for the supressed
$M$ band flux observed for much hotter objects (see \citet{NCE}
for an analysis of how the effects of non-equilibrium $\mathrm{CO}$
concentrations diminish with decreasing $\mathrm{T_{eff}}$).  
We note also that even if the supression of $M$ band flux remained, $M$ would
still be better for the detection than the $H$-regime.

The next apastron of $\epsilon$ Eri b is in 2010, and this
would be the best time to attempt to image it with a very
deep $M$ band observation.  We have observed that
our sensitivity in both speckle-limited and background-limited
regimes increases roughly as the square root of the integration
time, as we would expect.  Therefore, barring further
improvements in Clio or MMTAO, an exposure 35 (or $5.9^2$)
times as long as our $M$ band integration would be 
required to have a 40\% chance of making a confirmed
detection of the planet using Clio at the MMT.
This means 35 hours of observing, or about seven
good nights.  Improvements to the Clio instrument,
MMTAO, and our processing methods might bring the detection
in range with a shorter exposure, perhaps only two nights.
We note that at 1.7 arcseconds from $\epsilon$ Eri our
current images are speckle-limited -- the background
limit is still a factor of about 3 lower.  Alterations
to the instrument, or improved PSF subtraction methods 
in post-processing, may in future obtain near-background limited 
performance at this separation.  The planet might then 
be detectable with only one or two nights worth of integration,
though four to six nights would still be preferred to ensure
that an interesting upper limit could be set in the event of a non-detection.
Clio has already been used with a phase plate coronagraph \citep{phaseplate}
which improves the close-in sensitivity.   

As far as we know Clio, when used with the adaptive
secondary AO system of the MMT, is the only 
currently operating AO imager able to make the 
deep, high-efficiency integrations in the broad $M$ band required
to detect $\epsilon$ Eri b.  Other AO imagers exist that 
can use the $M'$ band, where the narrower bandpass reduces the
intensity of the thermal background.  However at $M'$ the sensitivity
to planets is also reduced and the project becomes unfeasible.  
Given a multi-night $M$ band integration with 
Clio, the goal of obtaining the first direct image of a 
mature extrasolar planet appears to be within reach.

Detection and characterization of $\epsilon$ Eri b should be
quite straightforward with new large telescopes such as the
LBT (which might be used instead of the MMT to make the first
detection), GMT, TMT, or E-ELT, provided the latter two are equipped
with the adaptive secondary AO systems necessary to reduce thermal
background and make deep $M$ band
observations feasible.  Space-based observations are likely to be
useful as well.  The planet might be studied at $L'$,
$M$ band, or longer wavelengths using JWST, or it could be detected
in reflected light at visible wavelengths by a sensitive space-based
coronagraph.  However, the first detection may come well ahead
of JWST and the next generation of giant telescopes --- it may
be achieved in the $M$ band with the MMT during the 2010 apastron.

\citet{ozernoy} and \citet{quillen} suggest that the dust disk
of $\epsilon$ Eri has been sculpted by a planet of 0.1-0.2 MJ in
an orbit between 40 and 65 AU in radius.  \citet{deller} agree,
and prefer the model of \citet{quillen}.  Such a planet
would be far too faint to detect with any telescope in the
near future.  However, \citet{deller} state
that an additional, $\sim$ 1 MJ planet in a closer-in orbit is
likely required to produce the observed clearing of the dust inside
of about 30 AU \citep{greaves}.  The RV/astrometric planet of \citet{hatzes} and
\citet{benedict} has too small an orbit to account for this dust clearing;
\citet{deller} suggest a larger orbital radius between 10 and 18 AU for the planet
responsible for clearing the dust.  \citet{benedict} mention a long-term 
trend in RV measurements for $\epsilon$ Eri A that might indicate just 
such a planet: a $\sim$ 1 MJ object
orbiting with a period longer than 50 years.  Since such a planet
would probably appear at least 3-4 arcsec from the star, we would likely
have detected it if it had a mass of 4-5 MJ or
greater, as would the \citet{GDPS} observation.  Since the mass
is expected to be closer to 1 MJ, it is not surprising the planet has not
yet been detected.  It might be imaged serendipitously in the course 
of a very long exposure intended to detect the known RV/astrometric planet.

\section{Conclusions} \label{conclusion}
We have taken very deep $L'$ and $M$ band images of the interesting
debris disk stars Vega and $\epsilon$ Eri to search each system
for orbiting planets and brown dwarfs.  For both stars we
obtained better sensitivity than shorter-wavelength observations
at small separations from the star.  The sensitivity of our observations
compared more favorably to the sensitivity of $H$-regime observations in the case
of $\epsilon$ Eri than in the case of Vega.  For $\epsilon$ Eri, our
$M$ band observation appears to set the best upper limits yet for
planets out to a separation of about 7 arcseconds, beyond which the
sensitivity of the \citet{GDPS} $H$-regime observations becomes very
slightly superior.


The reason our $\epsilon$ Eri observations have a
greater sensitivity advantage over $H$ regime observations than
do our images of Vega is the smaller distance to the $\epsilon$ Eri system.
This is another instance of the same physical reality we discussed
in Section \ref{finsens} above, when explaining why our $M$ band
sensitivity is much better than our $L'$ results on $\epsilon$ Eri
but not on Vega.  As we stated above, for $\epsilon$ Eri, 
the sensitivity of a given observation at any wavelength extends 
down to less luminous, lower $\mathrm{T_{eff}}$ planets than for 
Vega.  The $H-L'$ and $H-M$ colors, as well as the $L' - M$ color, of
low $\mathrm{T_{eff}}$ giant planets are more red than those of 
hotter ones.  Therefore the faintest detectable objects in the 
$\epsilon$ Eri system would be more red than those 
in the more distant Vega system, and longer wavelength observations 
are most useful for the nearer system.  This is a general and
important principle for planning optimal planet search strategies: 
the faintest detectable planets will be more red, and therefore
the relative advantage of long wavelengths over short ones will be higher, 
for the nearest stars.  For distant stars where only hot objects
with blue infrared colors can be detected, long wavelengths observations
are not as useful.  For very nearby stars such as $\epsilon$ Eri,
where very interesting, extremely low-mass, low $\mathrm{T_{eff}}$
planets can be detected, the long wavelengths are very useful because
the planets being sought have such red colors.

Planet-search observations at the $L'$ and $M$ bands have a considerable
advantage over those in the more commonly used $H$ band regime for
$\epsilon$ Eri and a handful of other bright, very nearby stars.  For
more distant bright stars such as Vega, $L'$ and $M$ band observations
give markedly better results only at separations inside about 3 arcsec,
and in this regime no currently employed method gives sensitivity
to any but the highest mass planets.  Observations in the bands we
have employed are still useful on Vega, but their use tends toward
a diversification of planet-search effort in case theoretical models
are overpredicting planets' $H$ band brightnesses.  For nearer
systems such as $\epsilon$ Eri, by contrast, $L'$ and $M$ band observations
clearly provide the best sensitivity at the most interesting separations,
and it is the $H$ regime images that naturally take the role of
diversifying effort under the supposition that the models may
overpredict planet brightness at longer wavelengths.

We have set a limit on the Vega planet hypothesized by \citet{wilner}
that is close to the 3 MJ mass he suggested for it.  It appears that
\citet{marois} could set a similar limit.  The evidence seems fairly
strong that no 3 MJ planet exists at this location.  This favors alternative
models involving smaller planets, such as those of \citet{marsh} and
\citet{wyatt}, or a 3 MJ planet in a larger orbit, such as that of
\citet{deller}.  Since a 3 MJ planet around Vega could be imaged in
multiple wavelength regimes with current technology, more 
submillimeter observations and 
further modeling to determine if such a planet is required to
explain the observed dust distribution is very desirable.  If this
does turn out to be the case, deep AO observations to detect the
planet could be strongly prioritized, and a sucessful detection in
one or more wavelength bands would be very likely.

Our $\epsilon$ Eri observation was not timed to catch the 
known planet $\epsilon$ Eri b at a large separation, and 
therefore our current data do not allow us to set an interesting 
limit on its mass.  \citet{janson} observed $\epsilon$ Eri at several 
epochs of more promising separation using SDI, and set limits 
in the 9-10 MJ range.

We have set a limit of 4-5 MJ for additional planets in 
more distant orbits around $\epsilon$ Eri.  The existence
of a planet in such an orbit may be indicated by a long term RV trend
\citep{benedict} and by a clearing of dust from the inner disk \citep{deller}. 
\citet{benedict} and \citet{deller} suggest a mass of around 1 MJ
for this hypothetical outer planet, so our non-detection is not
surprising.

We have explored the question of whether SDI imaging \citep{janson, biller} 
or $L'$ and $M$ band imaging (this work) is the method most 
likely ultimately to detect $\epsilon$ Eri b.
Our $M$ band images were much more sensitive at small separations than
our $L'$ results, so we have not considered the latter.  We find
that the sensitivity of the \citet{janson} observations
at the best epoch, where
the planet was near the optimal separation for SDI imaging, was
still insufficiently sensitive to detect the planet by a factor
of at least a thousand.  By contrast our observations, if carried 
out at apastron, would have missed the planet by a factor of only about 12.

This striking difference suggests that it is at $M$ band that the planet
$\epsilon$ Eri b will first be imaged.  A several-night observing 
campaign using Clio at the MMT might detect it during the 2010 
apastron passage, since we have observed the sensitivity in the 
speckle-dominated regions of $M$ band
images does go up approximately as the square root of the exposure
time.  More advanced PSF subtraction, or coronagraphic capability in
Clio \citep{phaseplate}, might reduce the required exposure time to detect the
planet to as little as 2 nights. 
At present, we believe Clio with MMTAO
is the only system capable of deep planet imaging
integrations in the $M$ band.  Spitzer, despite its enormously 
lower background and correspondingly excellent sensitivity,
does not have sufficient resolution to detect objects at the separations
expected for orbiting planets.

$\epsilon$ Eri b could be studied in more detail using
new giant telescopes such as the LBT and GMT with planned
adaptive secondary AO systems.  The $M$ band will remain
the best wavelength choice for observations using these
larger telescopes, so adaptive secondaries will remain
essential: conventional AO systems even on giant telescopes
will likely still have too high a thermal background for 
efficient, deep $M$ band images.  An $L'$ and $M$ band 
imager called LMIRCam is planned for the LBT \citep{elmer}.
When JWST is launched, it should also deliver interesting
scientific results on $\epsilon$ Eri b.  However, it is possible
that the first image of this planet --- the first direct
image of any mature extrasolar planet --- will be obtained using
Clio at the MMT in 2010.


\acknowledgments
This research has made use of the SIMBAD database,
operated at CDS, Strasbourg, France.

This research has made extensive use of information and code
from \citet{nrc}.

We thank I. Baraffe for kindly suppling us with theoretical
spectrum files corresponding to the planet models used in \citet{bar}.

We thank the referee, Remi Soummer, for helpful comments.

Facilities: \facility{MMT}

\end{document}